\documentclass[11pt]{article}
\usepackage{UF_FRED_paper_style}

\usepackage{longtable}
\usepackage{parskip}
\title{Evaluation of imputation techniques with varying percentage of missing data}

\author{Seema Sangari\\
    \href{mailto:ssangar1@students.kennesaw.edu}{\texttt{ssangar1@students.kennesaw.edu}} 
\and Herman E. Ray\\
    \href{mailto:hray8@kennesaw.edu}{\texttt{hray8@kennesaw.edu}} 
    }
    
\date{\today}

\begin{document}
{\setstretch{.8}
\maketitle
\begin{abstract}
Missing data is a common problem which has consistently plagued statisticians and applied analytical researchers.  While replacement methods like mean-based or hot deck imputation have been well researched, emerging imputation techniques enabled through improved computational resources have had limited formal assessment.  This study formally considers five more recently developed imputation methods -- \textit{Amelia, Mice, mi, Hmisc} and \textit{missForest} -- compares their performances using RMSE against actual values and against the well established mean-based replacement approach.  The RMSE measure was consolidated by method using a ranking approach.  Our results indicate that the \textit{missForest} algorithm performed best and the \textit{mi} algorithm performed worst. 

\noindent
\textit{\textbf{Keywords: }%
Missing Data; Missing at Random; Missing Completely at Random; Imputation; Multiple Imputation; MICE; mi; Amelia; missForest; Harrell Miscellaneous; RMSE} \\ 
\noindent

\end{abstract}
}


\section{Introduction}\label{s:Intro}
\noindent Missing data is a common problem encountered when analyzing data. \cite{Efron1994} defined missing data as a difficult problem due to the absence of some data elements in the familiar data structure. \cite{Graham2009} stated that the missing values significantly impact results drawn from the dataset. 
There are three major concerns related to missing data - reduction in statistical power, biased parameter estimation, and sample not being representative of the population.

\cite{Musil2002a} imputed missing at random data to compare five different approaches (list-wise deletion, mean substitution, simple regression,regression with an error term, and the expectation maximization [EM] algorithm) and compared the effects on descriptive statistics and correlation coefficients for imputed data and the complete data. They found mean substitution having the least impact whereas the regression with an error term and EM algorithm produced results closest to actual values.

 \cite{Baker2014} sought to find an appropriate imputation technique for health survey data and found that mean imputation was more accurate than multivariate normal and conditional autoregressive prior distribution-based imputation. \cite{Bono2007} investigated patient data and imputed with item-mean, pearson mean, regression, and hot-deck imputation. Unlike \cite{Musil2002a}, they found that all imputed values are comparable to the complete case mean values except regression imputation. 

On the other hand, \cite{Huisman2000} replaced values with different imputation techniques on 4 different datasets  and evaluated performance using Cronbach's alpha (\cite{Tatsuoka1971}) and Loevinger’s H-coefficient (\cite{Mokken1997}). They found that corrected variable mean (based on ability of the variable, i.e. score based on the observed values in the given variable when compared against the mean score of these values) substitution outperformed other imputation techniques. 

\cite{Barnes2006}  tested multiple imputation methods (regression-based MI methods, Bayesian least square (BLS), predictive mean matching (PMM), local random residual (LRR), modified propensity scores (MPS), completion score (CS)) and a last observation carried forward on small samples in clinical trials using simulation study and found BLS performed best and CS second best. 

\cite{Pantanowitz2009} assessed the impact of imputation on missing data with statistical analysis on classification problem. They imputed using random forests, auto-associative neural networks with genetic algorithms, auto-associative neuro-fuzzy configurations, and two random Forests and neural network. However, they did not find the impact of imputations significant.

\cite{Myrtveit2001} evaluated four imputation techniques (listwise deletion(LD), mean imputation (MeI), similar response pattern imputation (SRPI), and full information maximum likelihood (FIML)) from a software cost modeling perspective. They found that FIML performs better when data is not MCAR but the other techniques are biased with MCAR data. 

\cite{VanHulse2008} applied five imputation techniques (MeI, Regression Imputation (RI), Instance-based learning imputation (RI), REPTree imputation (RTI) and Bayesian multiple imputation (BMI)) on noisy software measurement data. They evaluated imputed values based on the impact of noise on imputation effectiveness using ANOVA and found that BMI and RI had the best results whereas MeI performed worst. 

\cite{Junninen2004} applied univariate (linear, spline, and nearest neighbor interpolation), multivariate (regression-based imputation (REGEM), nearest neighbor (NN), self-organizing map (SOM) and multi-layer perceptron (MLP)), and hybrid methods using historical simulated missing data patterns on air quality data and evaluated imputed values based on statistical measures - Index of agreement, $R^2$, RMSE, and Absolute MSE with bootstrapped standard errors. They found that the performance of multivariate methods can be improved slightly by using hybridization and more substantial multiple imputation where final imputed values are derived from various multivariate imputation results. 

\cite{Ambler2007} compared the imputation techniques and evaluated the imputed values using measure of agreement, rank correlation, RMSE, regression-based calibration measure and regression coefficients, and confidence interval coverage. Their experiment started with imputed values for the predictor variable. They found MICE outperformed with respect to the model estimation.

Most formal research to date has been done within a particular context e.g. model, software measurement, classification problem, etc. or application domain e.g. health, survey, clinical trials, quality measurement, etc. Table\ref{tab:LitSummary} provides the literature summary. The need to evaluate emerging imputation techniques in generalized contexts motivated this study. In this research, the objective is to evaluate the performance of six different imputation methods based on their imputed values against the original values independent of context. The experiment is designed in such a way that it started with the complete data set then different percentages, ranging from 5\% to 25\% in increments of 5\%, of the data were deleted randomly, and five different imputation techniques were used to impute the missing values. These imputed values were compared against the actual values using RMSE. Section \ref{s:MDIT} discusses categories of missing data and different imputation techniques used in this experiment. 
Section \ref{s:data} describes the data used for the study.  Section \ref{s:DoE} discusses the design of the experiment.  Section \ref{s:Results} discusses the findings of the experiment. Section \ref{s:Conclusion} provides conclusions based on the experiments. 

\begin{table}[ht!]
	\centering
	\caption{Literature Summary}
	\scalebox{.75}{
	\begin{tabular}{p{23em}p{15em}p{5em}}
	\hline
		\toprule
		\multicolumn{1}{l}{Techniques Evaluated} & \multicolumn{1}{l}{Key Findings} & Citations \\
		\midrule
		\hline \hline
		List-wise deletion, Mean imputation, simple regression, regression with an error term and expectation maximization algorithm &  
		- Better performance: Regression with an error term and EM algorithm  \newline
		- Worst Performance: mean Imputation
& \cite{Musil2002a} 
		\\ \\
		Mean Imputation, Multivariate Normal and conditional auto-regressive prior distribution & 

		- Better Performance: mean Imputation \newline
		- Lesson: Choose imputation based on the application

		 & \cite{Baker2014}  
		\\ \\
		Mean Imputation (Item-mean,Pearson mean), regression, and hot-deck imputation &
		- Better Performance: mean and Hot-deck imputation \newline
	    - Worse Performance: Regression based
		& \cite{Bono2007} 
		\\ \\
		Random draw substitution, Mean Imputation (item mean imputation, Pearson mean and corrected item mean imputation), item correlation imputation, hot-deck imputation  (hot-deck next case and  hot-deck nearest neighbor)& 	
		- Best Performance: Corrected variable mean

		&  \cite{Huisman2000}
			\\
		\\
		Regression-based Mean Imputation, Bayesian least square(BLS), predictive mean matching(PMM), local random residual(LRR), modified propensity scores(MPS), completion score(CS)) and a last observation carried forward on small samples & 
		- Best Performance: BLS and CS
		& \cite{Barnes2006} 
		\\
		\\
		Random forests, auto-associative neural networks with genetic algorithms, auto-associative neuro-fuzzy configurations, and two random forests and neural network & 
	    - Imputation didn't give significant results 
		& \cite{Pantanowitz2009} \\
		\\
		Listwise deletion(LD), mean imputation , similar response pattern imputation (SRPI), and full information maximum likelihood (FIML) 
		&
		- FIML performs better with non MCAR data \newline
		- Other techniques: Biased with MCAR data
		  
		  & 
		  \cite{Myrtveit2001}  
		  \\ \\
		  
		Mean Imputation, Regression Imputation (RI), Instance-based learning imputation (RI), REPTree imputation (RTI) and Bayesian multiple imputation(BMI) 
		& 
		- Better Performance: BMI and BI \newline
		- Worse Performance: Mean Imputation
        & \cite{VanHulse2008} 
		\\ \\
		Univariate (linear, spline, and nearest neighbor interpolation), multivariate (regression-based imputation (REGEM), nearest neighbor (NN), self-organizing map (SOM) and multi-layer perceptron (MLP)), and hybrid methods using historical simulated missing data & 
		- Lesson: Hybridization improves performance of multivariate methods 
		& \cite{Junninen2004} \\
	\\
		Mean imputation, Conditional mean imputation, Multiple Imputations - Hotdecking, Hotdecking by covariate pattern, Hotdecking by observation, Hotdecking including outcome, Multiple imputation by Chained Equation (MICE) &  

		- Best Performance: MICE with least biased estimate \newline
		- Better Performance: Conditional mean imputation but inappropriate for variable selection methods
		& \cite{Ambler2007} 
		\\	
		\bottomrule
		\hline
	\end{tabular}%
	}
	\label{tab:LitSummary}%
\end{table}%

\section{Theoretical Concepts }\label{s:MDIT}
In this study, Missing Completely at Random missing data  is considered. Six different imputation methods were used to impute values including single-value imputation using the mean. 
The six techniques evaluated here are \textit{mean imputation}, \textit{multiple imputation by chained equation}, \textit{multiple imputation with diagnostics}, \textit{amelia}, \textit{harrell miscellaneous}, and \textit{missForest}. The three assumptions of missingness  are \textit{missing at random}, \textit{missing completely at random}, and \textit{missing not at random}.

\subsection{Missing Data}\label{s:MissingData}
\cite{Efron1994} defined missing data as a problem caused by the absence of a familiar data structure. \cite{Rubin1976} classified missing data into three classes based on the likelihood of missing.
\begin{itemize}
	\item 
	Missing At Random (MAR) – Missing probability may depend on observed data but not on unobserved data (\cite{Little2002}), i.e. systematically related to observed data and not to unobserved data. \cite{VanBuuren2012} defined MAR data as when the probability of missing is the same within observed data variables. 

	\item 
	Missing Completely at Random (MCAR) - Missing probability is independent of observed and unobserved data (\cite{Little2002}). It is considered special case of MAR when there is no systematic difference between the variables with missing data and variables with complete data. \cite{VanBuuren2012} defined MCAR data when the probability of missing is the same for all variables. 
	
	\item 
	Missing Not At Random (MNAR)- Missing probability does not depend on unobserved data. Mack et al defined MNAR as when missing data is not related to any measurable events or factors (\cite{Mack2018}). \cite{VanBuuren2012} defined MNAR data as when it is neither MAR nor MCAR data.
\end{itemize}

\subsection{Imputation Techniques }\label{s:Imputations}
Imputation is the process of replacing missing values with appropriate meaningful estimates. Researchers are frequently tempted to delete the observations or variables with missing values, however, this has potential to lead to information loss impacting results. Another consideration is also pairwise deletion in which the analysis is done with complete cases of relevant variables; as a result, the sample size differs for different dependent variables. Six different imputations are investigated, discussed as below:
\subsubsection{Mean Imputation}
	 Missing values are replaced with the column-based mean i.e. individual variable mean value. This method is the easiest but not very accurate. It does not take into account the correlation between the explanatory variables. One may also consider median or mode substitution if the data is skewed.
\subsubsection{Multiple Imputation by Chained Equation (MICE)} MICE works in iterations in which imputations are done for each variable one by one. \cite{Azur2011} defined the approach - First, initiate by replacing all missing values with individual predictor variables, referring to mean imputation as ``place holders". Second, to impute a variable, $v$,mean values are replaced back missing values, variable $v$ is now a dependent variable and is regressed over the other variables working as independent variables, missing values are then replaced by the predictions based on the regression model. Third, To impute other missing variables, imputed values of variable $v$ where $v$ acts as independent variable whereas rest of the variables will be used with mean substitution. Fourth, second \& third steps are repeated until each variable is imputed with predictions.
	These steps are one iteration or cycle. The number of iterations/cycles are repeated, and imputations are updated.
	\subsubsection{Multiple Imputation with Diagnostics (mi)}
	\cite{Su2011} stated that \textit{mi} imputation technique is derived from MICE but with one of the key differences that it imputes from conditional distribution of a variable whereas other variables are either imputed or observed. The benefit of mi over MICE is that it has the capability to deal with irregularities in the data, e.g. multi-collinearity\footnote{A problem where two or more predictors are highly linearly associated} within a dataset. To impute a variable, the procedure is split into four steps (\cite{Su2011})-
First, setup analyzes missing data patterns to recognize problems in data structure, performs preprocessing and identifies conditional models.
Second, iterates over MICE based imputations but with a conditional model, and tests imputed values for conditionality, acceptability, and  convergence.
Third, analysis obtains multiple imputed complete datasets and  pools them for the complete case analysis.
Fourth, validation analyzes sensitivity, performs cross-validation, and tests for compatibility.
	\subsubsection{Amelia}
	\cite{Honaker2011} mentioned that the algorithm assumes complete data that follows a multivariate normal distribution. The algorithm works well if the data is sampled from other distributions (\cite{Schafer1997,Schafer1999a}). It imputes based on a Bootstrapping and Expectation-Maximization (EMB) algorithm. 
First, EMB algorithm draws from the posterior by integrating the EM algorithm with bootstrapping where each bootstrapped sample introduces uncertainty and the EM algorithm finds the mode of the posterior (\cite{Dempster1977}), which allows for fundamental uncertainty(\cite{Honaker2010}).
Second, EMB is applied to generate parameters for the complete data. 
Third, imputes with the values drawn from conditional distribution on observed and complete data parameters, $\theta = (\mu,\Sigma)$ - linear regression with parameters (\cite{Honaker2011}).
Fourth, pools the multiple imputations using an average of the estimates.
	\subsubsection{Harrell Miscellaneous (Hmisc)}
	\cite{R-Package2010}, and  \cite{Harrell2016} stated that the algorithm works at simple imputation based on mean/mode/median as well as multiple imputation based on additive regression, bootstrapping, and predictive mean matching approaches. To impute missing values in each variable,
First,i nitiate the missing values from randomized sample of non-missing values of size `m'.
Second, fit the flexible additive model to find the optimum transformation. Identity transformation can be forced as well.
Third, apply the flexible fitted model to make predictions for the non-missing observed values.
Fourth, impute the missing value with the observed value, where the predicted transformed value is nearest to the predicted transformed value of the missing value.
Fifth, to impute other variables, randomly draw the imputed values of the variable.
	For n iterations, the first set of 'x' iterations are the burn-in set.
	\subsubsection{missForest }
	\cite{Stekhoven2012} defined this as a non-parametric approach in which variables are pairwise independent. The algorithm is based on the random forest approach (\cite{Breiman2001}). For each variable, \textit{missForest}  generates random forest with the observed values and predicts to impute missing values. The algorithm repeats itself until the number of iterations is maximized or stopping criterion is met.  \cite{Oshiro2012}  recommended forest with   trees between 64 and 128.
	
	\subsection{Computation}
	In terms of computation, single-value imputation i.e. \textit{mean} imputation is the fastest algorithm. The algorithms \textit{Amelia}, \textit{mi} and \text{Hmisc}  imputed quickly. \textit{MICE} algorithm relatively more time to impute. None of these techniques were computationally demanding. However, \textit{missForest} algorithm is computationally intensive and time consuming. For the purpose of this experiment, minimum recommended number of trees, 64 trees, were generated in forest (\cite{Oshiro2012}). The \textit{missForest} algorithm  run on more than 20 cores on GPU in parallel. Despite using the GPU and 20 cores, the single run of one iteration took more than 45 minutes to impute whereas rest of the algorithms run five iterations in couple of minutes on single core to impute.

\section{Data}\label{s:data}
The current study uses data provided by a major U.S. Credit Bureau and includes consumer credit utilization data from 2008, with all personally identifiable fields omitted prior to receipt. The multiple data sets were appropriately merged resulting in 1.2 million records with 152 input variables and one binary dependent variable, deciding factor whether credit should be given or not. Missing values are part of the data: columns with more than 40\% of missing data were removed from the analysis.  For the purpose of this study, only complete records were considered. The data set was a reduced to almost a quarter of the data - 329,917 records with 153 variables. Finally, the data was reduced to 329,917 records with the 21 most prominent variables using LASSO with hyper-parameter $\lambda$ chosen to be 0.01 using 10-fold cross-validation. For the analysis, only 21 selected variables were used. All variables  are continuous.

\section{Experiment Design}\label{s:DoE}
The experiment is designed to discover the impact of the percent of missing data on the choice of imputation algorithm. The experiment is conducted on a data set with 329,917 records with 21 predictor variables. The experiment is designed such that 
	50 random samples of 50,000 records were generated from the data.
	Percentage of the data is deleted from each variable using the MCAR\footnote{Amelia Algorithm assumes the missingness to MAR, but there is no way to check if the missing data is MAR. Therefore, in order to ensure MAR -special case of MAR applied, MCAR}$^{,}$\footnote{MCAR behavior in the data can be tested by turning the dataset into a binary dummy dataset where missing value takes 1 otherwise 0 and applying Welsh's t-test on each pair of variables in the dummy dataset.} approach - 5\%, 10\%, 15\%, 20\%, and 25\%.
	Each sample is standardized using the mean and standard deviation from the data with missing values (ignoring missing value).
		Standardization is done on the random sample after the MCAR process is applied. 
		 For calculation of RMSE, the actual values are also standardized with parameters based on the data with missing values.
	Missing values are imputed based on six different approaches - mean, MICE, mi, Amelia, Hmisc, and missForest.
	 For each pair of missing percentage and imputation method, 5 iterations were computed apart from mean as it will remain constant for each iteration.
	RMSE measure of difference between the imputed values and actual values is used to compare the various imputation approaches.

\section{Results}\label{s:Results}
The study showed that the percentage of missing data does not impact the choice of algorithm. For each imputation technique, RMSE was computed to analyze the performance of each iteration with respect to algorithm and missing percentage. 

\begin{figure} 
		\raggedleft
		\addtocounter{subfigure}{-1}
		\subfigure{ \frame{\includegraphics[width=0.33\textwidth ]{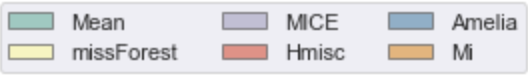}}}
		\begin{center}
		\subfigure[Mean substitution under-performs]{%
			\label{fig:Underperforms}
			\frame{\includegraphics[width=0.33\textwidth]{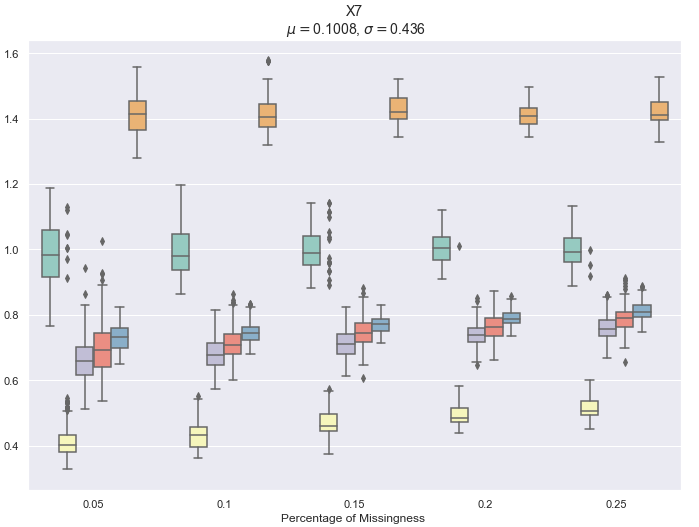}}
		}%
		\subfigure[Mean substitution performs at par]{%
			\label{fig:AtPar}
			\frame{\includegraphics[width=0.33\textwidth]{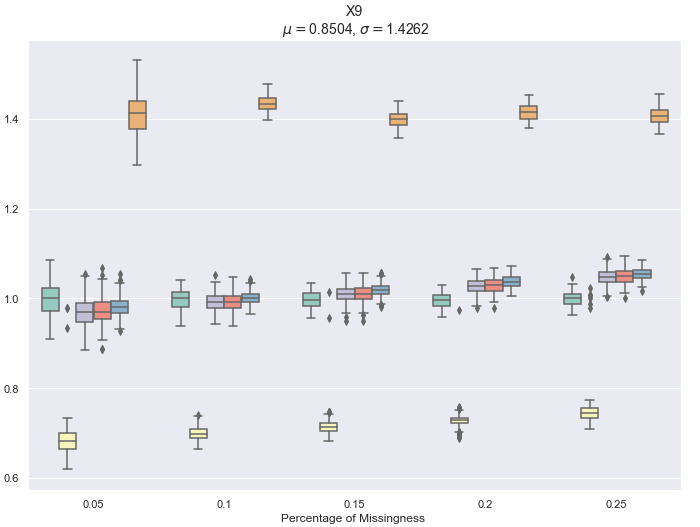}}
		}%
		\subfigure[Mean substitution out-performs]{%
			\label{fig:OutPerforms}
			\frame{\includegraphics[width=0.33\textwidth]{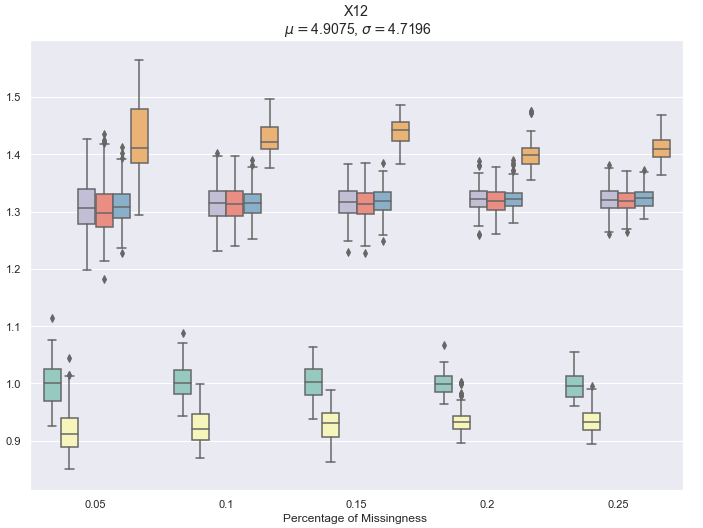}}
		}%
		\end{center}
		\caption{BoxPlots based on RMSE Results}%
		\label{fig:MeanPerformanceBoxPlots}
\end{figure}

To visualize the results, box-plots were generated for each variable to show the performance of each imputation algorithm corresponding to different missing percentages. Figure \ref{fig:MeanPerformanceBoxPlots} shows three of them to illustrate results
	The best performing algorithm was found to be \textit{missForest}. However, it is computationally intensive and time consuming. The algorithm consistently gave the lowest RMSE in each case - implying the imputations were closest with \textit{missForest} algorithm.
	The worst performing algorithm was found to be \textit{mi}. The algorithm consistently gave the highest RMSE in each case - implying the imputations were farthest with \textit{mi} algorithm.
	Three algorithms, \textit{MICE, Amelia and Hmisc}, performed with minimal differences, indicating the imputations based on either of these approaches would not be far from each other. 
Always performed better than \textit{mi} algorithm but worse than \textit{missForest}. 
 One of the three algorithms can be chosen for analysis as their results are not far from each other.	
	 Mean substitution either under-performed(Fig-\ref{fig:Underperforms} or out-performed(Fig-\ref{fig:OutPerforms}) or was at par(Fig-\ref{fig:AtPar} with  \textit{MICE, Amelia, and Hmisc} algorithms.
	Mean substitution outperforms for 14 variables.
	 Mean substitution underperforms for six variables.
	Mean substitution performs almost at par for one variable.
The boxplots of other variables are shown in Appendix \ref{ssec:RMSEBoxPlots}.
%

In order to further explore the imputed values, an ANOVA test was applied to test if the imputations using MICE, Amelia, and Hmisc are statistically similar for each of 21 variables.
	\textit{MICE}, \textit{Amelia}, and \textit{Hmisc} imputations are found to be statistically similar for four out of 21 variables. 
	 \textit{MICE} and \textit{Amelia} imputations are found to be statistically similar for five out of 21 variables.
	\textit{MICE} and \textit{Hmisc} imputations are found to be statistically similar for 11 out of 21 variables.
	\textit{Amelia} and \textit{Hmisc} imputations are found to be statistically similar for five out of 21 variables.

For each variable, the multiple measures -- mean, median, Q1, Q2, Q3, Q4 and standard deviation -- were computed. These RMSE measures for six imputation techniques --  computed for 5\%, 15\%, and 25\% -- given in Appendix \ref{ssec:RMSEResults}. 
Ranks for mean and standard deviation are assigned to each imputation technique ranging from one to six for each variable where Rank one indicates best and six indicates worst. These results are further consolidated based on these ranks. For each imputation technique, ranks were consolidated in Table \ref{tab:ConsolidatedResults} using average and median. The table shows that the \textit{missForest} algorithm performs best and the \textit{Mi} algorithm performs worst w.r.t. average and median of ranks of algorithms against  RMSE values despite the wide range of its standard deviation indicated by higher rank of average and median of standard deviation.

 \begin{table}
 	\centering
 	\setlength\tabcolsep{0pt}
 	\caption{Results consolidated based on Ranks}
 	\scriptsize
 	 \scalebox{1}{
 	 \begin{tabular*}{\textwidth}{@{\extracolsep{\fill}}*{13}{r}}
 		\hline
 		& \multicolumn{4}{c}{\textbf{5\% Missing Data} }                             & \multicolumn{4}{c}{\textbf{15\% Missing Data} }                          & \multicolumn{4}{c} {\textbf{25\% Missing Data}}                             \\ 
 		\cline{2-5}  \cline{6-9} \cline{10-13}
 		& \multicolumn{2}{c}{Mean} & \multicolumn{2}{c}{Standard Deviation} & \multicolumn{2}{c}{Mean} & \multicolumn{2}{c}{Standard Deviation} & \multicolumn{2}{c}{Mean} & \multicolumn{2}{c}{Standard Deviation}  \\ 
 			\cline{2-3}  \cline{4-5} \cline{6-7} \cline{8-9} \cline{10-11} \cline{12-13}
 		Method/Rank & Average & Median         & Average & Median                       & Average & Median         & Average & Median                       & Average & Median         & Average & Median                        \\
 		\hline \hline
 		Mean        & 4       & 5              & 4       & 4                            & 4       & 5              & 4       & 5                            & 4       & 5              & 4       & 4                             \\
 		MICE        & 3       & 2              & 3       & 2                            & 3       & 3              & 3       & 3                            & 3       & 3              & 3       & 2                             \\
 		Mi          & 6       & 6              & 6       & 5                            & 6       & 6              & 3       & 4                            & 6       & 6              & 4       & 4                             \\
 		Amelia      & 4       & 4              & 4       & 1                            & 4       & 4              & 2       & 1                            & 4       & 4              & 2       & 1                             \\
 		Hmisc       & 3       & 3              & 3       & 3                            & 3       & 3              & 4       & 3                            & 3       & 3              & 4       & 3                             \\
 		missForest  & 1       & 1              & 1       & 6                            & 1       & 1              & 5       & 6                            & 1       & 1              & 5       & 6                             \\
 		\hline
	\end{tabular*}
}
	\label{tab:ConsolidatedResults}%
 \end{table}

\section{Conclusion}\label{s:Conclusion}
The study was conducted to determine if there is an impact on choice of algorithm with change in missing percentage in the dataset. In this study, 50 random samples with 50,000 records were collected from a complete dataset with more than 300,000 records. The missing data was generated using the MCAR approach with varying percentages of missingness - 5\%, 10\%, 15\%, 20\%, and 25\%. The missing data was imputed with five different algorithms in addition to mean-based imputation; five iterations were computed for each algorithm and missing percentage of missingness. The results showed that there is no impact on choice of algorithm with change in missing percentage. Based on RMSE, the study showed that \textit{missForest} algorithm performed best whereas \textit{mi} algorithm performed wost irrespective of the missing percentage. The performance of \textit{MICE}, \textit{Amelia}, and \textit{Hmisc} was close to each other - RMSE of \textit{MICE} \& \textit{Hmisc} are statistically similar for a little more than 50\% of variables, and RMSE of \textit{MICE} \& \textit{Amelia} and  \textit{Amelia} \& \textit{Hmisc} are statistically similar for 25\% of variables  based on ANOVA test results. These three algorithms and \textit{mean} imputations always performed better than the \textit{mi} algorithm but worse than the \textit{missForest} algorithm. 

\medskip

\bibliography{ImputationTechniques.bib}

\begin{thebibliography}{}

\bibitem [\protect \citeauthoryear {%
Ambler%
, Omar%
\BCBL {}\ \BBA {} Royston%
}{%
Ambler%
\ \protect \BOthers {.}}{%
{\protect \APACyear {2007}}%
}]{%
Ambler2007}
\APACinsertmetastar {%
Ambler2007}%
\begin{APACrefauthors}%
Ambler, G.%
, Omar, R\BPBI Z.%
\BCBL {}\ \BBA {} Royston, P.%
\end{APACrefauthors}%
\unskip\
\newblock
\APACrefYearMonthDay{2007}{}{}.
\newblock
{\BBOQ}\APACrefatitle {{A comparison of imputation techniques for handling
  missing predictor values in a risk model with a binary outcome}} {{A
  comparison of imputation techniques for handling missing predictor values in
  a risk model with a binary outcome}}.{\BBCQ}
\newblock
\APACjournalVolNumPages{Statistical Methods in Medical Research}{}{}{}.
\newblock
\begin{APACrefDOI} \doi{10.1177/0962280206074466} \end{APACrefDOI}
\PrintBackRefs{\CurrentBib}

\bibitem [\protect \citeauthoryear {%
Azur%
, Stuart%
, Frangakis%
\BCBL {}\ \BBA {} Leaf%
}{%
Azur%
\ \protect \BOthers {.}}{%
{\protect \APACyear {2011}}%
}]{%
Azur2011}
\APACinsertmetastar {%
Azur2011}%
\begin{APACrefauthors}%
Azur, M\BPBI J.%
, Stuart, E\BPBI A.%
, Frangakis, C.%
\BCBL {}\ \BBA {} Leaf, P\BPBI J.%
\end{APACrefauthors}%
\unskip\
\newblock
\APACrefYearMonthDay{2011}{}{}.
\newblock
{\BBOQ}\APACrefatitle {{Multiple imputation by chained equations: What is it
  and how does it work?}} {{Multiple imputation by chained equations: What is
  it and how does it work?}}{\BBCQ}
\newblock
\APACjournalVolNumPages{International Journal of Methods in Psychiatric
  Research}{}{}{}.
\newblock
\begin{APACrefDOI} \doi{10.1002/mpr.329} \end{APACrefDOI}
\PrintBackRefs{\CurrentBib}

\bibitem [\protect \citeauthoryear {%
Baker%
, White%
\BCBL {}\ \BBA {} Mengersen%
}{%
Baker%
\ \protect \BOthers {.}}{%
{\protect \APACyear {2014}}%
}]{%
Baker2014}
\APACinsertmetastar {%
Baker2014}%
\begin{APACrefauthors}%
Baker, J.%
, White, N.%
\BCBL {}\ \BBA {} Mengersen, K.%
\end{APACrefauthors}%
\unskip\
\newblock
\APACrefYearMonthDay{2014}{}{}.
\newblock
{\BBOQ}\APACrefatitle {{Missing in space: An evaluation of imputation methods
  for missing data in spatial analysis of risk factors for type II diabetes}}
  {{Missing in space: An evaluation of imputation methods for missing data in
  spatial analysis of risk factors for type II diabetes}}.{\BBCQ}
\newblock
\APACjournalVolNumPages{International Journal of Health Geographics}{}{}{}.
\newblock
\begin{APACrefDOI} \doi{10.1186/1476-072X-13-47} \end{APACrefDOI}
\PrintBackRefs{\CurrentBib}

\bibitem [\protect \citeauthoryear {%
Barnes%
, Lindborg%
\BCBL {}\ \BBA {} Seaman%
}{%
Barnes%
\ \protect \BOthers {.}}{%
{\protect \APACyear {2006}}%
}]{%
Barnes2006}
\APACinsertmetastar {%
Barnes2006}%
\begin{APACrefauthors}%
Barnes, S\BPBI A.%
, Lindborg, S\BPBI R.%
\BCBL {}\ \BBA {} Seaman, J\BPBI W.%
\end{APACrefauthors}%
\unskip\
\newblock
\APACrefYearMonthDay{2006}{}{}.
\newblock
{\BBOQ}\APACrefatitle {{Multiple imputation techniques in small sample clinical
  trials}} {{Multiple imputation techniques in small sample clinical
  trials}}.{\BBCQ}
\newblock
\APACjournalVolNumPages{Statistics in Medicine}{}{}{}.
\newblock
\begin{APACrefDOI} \doi{10.1002/sim.2231} \end{APACrefDOI}
\PrintBackRefs{\CurrentBib}

\bibitem [\protect \citeauthoryear {%
Bono%
, Ried%
, Kimberlin%
\BCBL {}\ \BBA {} Vogel%
}{%
Bono%
\ \protect \BOthers {.}}{%
{\protect \APACyear {2007}}%
}]{%
Bono2007}
\APACinsertmetastar {%
Bono2007}%
\begin{APACrefauthors}%
Bono, C.%
, Ried, L\BPBI D.%
, Kimberlin, C.%
\BCBL {}\ \BBA {} Vogel, B.%
\end{APACrefauthors}%
\unskip\
\newblock
\APACrefYearMonthDay{2007}{}{}.
\newblock
{\BBOQ}\APACrefatitle {{Missing data on the Center for Epidemiologic Studies
  Depression Scale: A comparison of 4 imputation techniques}} {{Missing data on
  the Center for Epidemiologic Studies Depression Scale: A comparison of 4
  imputation techniques}}.{\BBCQ}
\newblock
\APACjournalVolNumPages{Research in Social and Administrative Pharmacy}{}{}{}.
\newblock
\begin{APACrefDOI} \doi{10.1016/j.sapharm.2006.04.001} \end{APACrefDOI}
\PrintBackRefs{\CurrentBib}

\bibitem [\protect \citeauthoryear {%
Breiman%
}{%
Breiman%
}{%
{\protect \APACyear {2001}}%
}]{%
Breiman2001}
\APACinsertmetastar {%
Breiman2001}%
\begin{APACrefauthors}%
Breiman, L.%
\end{APACrefauthors}%
\unskip\
\newblock
\APACrefYearMonthDay{2001}{}{}.
\newblock
{\BBOQ}\APACrefatitle {{Random forests}} {{Random forests}}.{\BBCQ}
\newblock
\APACjournalVolNumPages{Machine Learning}{}{}{}.
\newblock
\begin{APACrefDOI} \doi{10.1023/A:1010933404324} \end{APACrefDOI}
\PrintBackRefs{\CurrentBib}

\bibitem [\protect \citeauthoryear {%
Dempster%
, Laird%
\BCBL {}\ \BBA {} Rubin%
}{%
Dempster%
\ \protect \BOthers {.}}{%
{\protect \APACyear {1977}}%
}]{%
Dempster1977}
\APACinsertmetastar {%
Dempster1977}%
\begin{APACrefauthors}%
Dempster, A\BPBI P.%
, Laird, N\BPBI M.%
\BCBL {}\ \BBA {} Rubin, D\BPBI B.%
\end{APACrefauthors}%
\unskip\
\newblock
\APACrefYearMonthDay{1977}{}{}.
\newblock
{\BBOQ}\APACrefatitle {{ Maximum Likelihood from Incomplete Data Via the EM
  Algorithm }} {{ Maximum Likelihood from Incomplete Data Via the EM Algorithm
  }}.{\BBCQ}
\newblock
\APACjournalVolNumPages{Journal of the Royal Statistical Society: Series B
  (Methodological)}{}{}{}.
\newblock
\begin{APACrefDOI} \doi{10.1111/j.2517-6161.1977.tb01600.x} \end{APACrefDOI}
\PrintBackRefs{\CurrentBib}

\bibitem [\protect \citeauthoryear {%
Efron%
}{%
Efron%
}{%
{\protect \APACyear {1994}}%
}]{%
Efron1994}
\APACinsertmetastar {%
Efron1994}%
\begin{APACrefauthors}%
Efron, B.%
\end{APACrefauthors}%
\unskip\
\newblock
\APACrefYearMonthDay{1994}{}{}.
\newblock
{\BBOQ}\APACrefatitle {{Missing data, imputation, and the bootstrap}} {{Missing
  data, imputation, and the bootstrap}}.{\BBCQ}
\newblock
\APACjournalVolNumPages{Journal of the American Statistical Association}{}{}{}.
\newblock
\begin{APACrefDOI} \doi{10.1080/01621459.1994.10476768} \end{APACrefDOI}
\PrintBackRefs{\CurrentBib}

\bibitem [\protect \citeauthoryear {%
{Frank E Harrell Jr}%
}{%
{Frank E Harrell Jr}%
}{%
{\protect \APACyear {2010}}%
}]{%
R-Package2010}
\APACinsertmetastar {%
R-Package2010}%
\begin{APACrefauthors}%
{Frank E Harrell Jr}.%
\end{APACrefauthors}%
\unskip\
\newblock
\APACrefYearMonthDay{2010}{}{}.
\newblock
{\BBOQ}\APACrefatitle {{Hmisc}} {{Hmisc}}.{\BBCQ}
\newblock
\APACjournalVolNumPages{R-Package}{}{}{}.
\newblock
\begin{APACrefURL} \url{http://biostat.mc.vanderbilt.edu/Hmisc}
  \end{APACrefURL}
\PrintBackRefs{\CurrentBib}

\bibitem [\protect \citeauthoryear {%
Graham%
}{%
Graham%
}{%
{\protect \APACyear {2009}}%
}]{%
Graham2009}
\APACinsertmetastar {%
Graham2009}%
\begin{APACrefauthors}%
Graham, J\BPBI W.%
\end{APACrefauthors}%
\unskip\
\newblock
\APACrefYearMonthDay{2009}{}{}.
\newblock
{\BBOQ}\APACrefatitle {{Missing Data Analysis: Making It Work in the Real
  World}} {{Missing Data Analysis: Making It Work in the Real World}}.{\BBCQ}
\newblock
\APACjournalVolNumPages{Annual Review of Psychology}{}{}{}.
\newblock
\begin{APACrefDOI} \doi{10.1146/annurev.psych.58.110405.085530}
  \end{APACrefDOI}
\PrintBackRefs{\CurrentBib}

\bibitem [\protect \citeauthoryear {%
Harrell%
\ \BBA {} Dupont%
}{%
Harrell%
\ \BBA {} Dupont%
}{%
{\protect \APACyear {2016}}%
}]{%
Harrell2016}
\APACinsertmetastar {%
Harrell2016}%
\begin{APACrefauthors}%
Harrell, F\BPBI E.%
\BCBT {}\ \BBA {} Dupont, C.%
\end{APACrefauthors}%
\unskip\
\newblock
\APACrefYearMonthDay{2016}{}{}.
\newblock
{\BBOQ}\APACrefatitle {{Package ‘Hmisc': Harrell Miscellaneous}} {{Package
  ‘Hmisc': Harrell Miscellaneous}}.{\BBCQ}
\newblock
\APACjournalVolNumPages{R topics Documented}{}{}{}.
\PrintBackRefs{\CurrentBib}

\bibitem [\protect \citeauthoryear {%
Honaker%
\ \BBA {} King%
}{%
Honaker%
\ \BBA {} King%
}{%
{\protect \APACyear {2010}}%
}]{%
Honaker2010}
\APACinsertmetastar {%
Honaker2010}%
\begin{APACrefauthors}%
Honaker, J.%
\BCBT {}\ \BBA {} King, G.%
\end{APACrefauthors}%
\unskip\
\newblock
\APACrefYearMonthDay{2010}{}{}.
\newblock
{\BBOQ}\APACrefatitle {{What to do about missing values in time-series
  cross-section data}} {{What to do about missing values in time-series
  cross-section data}}.{\BBCQ}
\newblock
\APACjournalVolNumPages{American Journal of Political Science}{}{}{}.
\newblock
\begin{APACrefDOI} \doi{10.1111/j.1540-5907.2010.00447.x} \end{APACrefDOI}
\PrintBackRefs{\CurrentBib}

\bibitem [\protect \citeauthoryear {%
Honaker%
, King%
\BCBL {}\ \BBA {} Blackwell%
}{%
Honaker%
\ \protect \BOthers {.}}{%
{\protect \APACyear {2011}}%
}]{%
Honaker2011}
\APACinsertmetastar {%
Honaker2011}%
\begin{APACrefauthors}%
Honaker, J.%
, King, G.%
\BCBL {}\ \BBA {} Blackwell, M.%
\end{APACrefauthors}%
\unskip\
\newblock
\APACrefYearMonthDay{2011}{}{}.
\newblock
{\BBOQ}\APACrefatitle {{Amelia II: A program for missing data}} {{Amelia II: A
  program for missing data}}.{\BBCQ}
\newblock
\APACjournalVolNumPages{Journal of Statistical Software}{}{}{}.
\newblock
\begin{APACrefDOI} \doi{10.18637/jss.v045.i07} \end{APACrefDOI}
\PrintBackRefs{\CurrentBib}

\bibitem [\protect \citeauthoryear {%
Huisman%
}{%
Huisman%
}{%
{\protect \APACyear {2000}}%
}]{%
Huisman2000}
\APACinsertmetastar {%
Huisman2000}%
\begin{APACrefauthors}%
Huisman, M.%
\end{APACrefauthors}%
\unskip\
\newblock
\APACrefYearMonthDay{2000}{}{}.
\newblock
{\BBOQ}\APACrefatitle {{Imputation of missing item responses: Some simple
  techniques}} {{Imputation of missing item responses: Some simple
  techniques}}.{\BBCQ}
\newblock
\APACjournalVolNumPages{Quality and Quantity}{}{}{}.
\newblock
\begin{APACrefDOI} \doi{10.1023/A:1004782230065} \end{APACrefDOI}
\PrintBackRefs{\CurrentBib}

\bibitem [\protect \citeauthoryear {%
Junninen%
, Niska%
, Tuppurainen%
, Ruuskanen%
\BCBL {}\ \BBA {} Kolehmainen%
}{%
Junninen%
\ \protect \BOthers {.}}{%
{\protect \APACyear {2004}}%
}]{%
Junninen2004}
\APACinsertmetastar {%
Junninen2004}%
\begin{APACrefauthors}%
Junninen, H.%
, Niska, H.%
, Tuppurainen, K.%
, Ruuskanen, J.%
\BCBL {}\ \BBA {} Kolehmainen, M.%
\end{APACrefauthors}%
\unskip\
\newblock
\APACrefYearMonthDay{2004}{}{}.
\newblock
{\BBOQ}\APACrefatitle {{Methods for imputation of missing values in air quality
  data sets}} {{Methods for imputation of missing values in air quality data
  sets}}.{\BBCQ}
\newblock
\APACjournalVolNumPages{Atmospheric Environment}{}{}{}.
\newblock
\begin{APACrefDOI} \doi{10.1016/j.atmosenv.2004.02.026} \end{APACrefDOI}
\PrintBackRefs{\CurrentBib}

\bibitem [\protect \citeauthoryear {%
Little%
\ \BBA {} Rubin%
}{%
Little%
\ \BBA {} Rubin%
}{%
{\protect \APACyear {2002}}%
}]{%
Little2002}
\APACinsertmetastar {%
Little2002}%
\begin{APACrefauthors}%
Little, R\BPBI J\BPBI A.%
\BCBT {}\ \BBA {} Rubin, D\BPBI B.%
\end{APACrefauthors}%
\unskip\
\newblock
\APACrefYear{2002}.
\newblock
\APACrefbtitle {{Statistical Analysis with Missing Data}} {{Statistical
  Analysis with Missing Data}}\ (\PrintOrdinal{Second}\ \BEd).
\newblock
\APACaddressPublisher{}{Wiley-Interscience;}.
\newblock
\begin{APACrefDOI} \doi{10.1002/9781119013563} \end{APACrefDOI}
\PrintBackRefs{\CurrentBib}

\bibitem [\protect \citeauthoryear {%
Mack%
, Su%
\BCBL {}\ \BBA {} Westreich%
}{%
Mack%
\ \protect \BOthers {.}}{%
{\protect \APACyear {2018}}%
}]{%
Mack2018}
\APACinsertmetastar {%
Mack2018}%
\begin{APACrefauthors}%
Mack, C.%
, Su, Z.%
\BCBL {}\ \BBA {} Westreich, D.%
\end{APACrefauthors}%
\unskip\
\newblock
\APACrefYear{2018}.
\newblock
\APACrefbtitle {{Managing Missing Data in Patient Registries}} {{Managing
  Missing Data in Patient Registries}}.
\newblock
\APACaddressPublisher{}{Agency for Healthcare Research and Quality (US)}.
\PrintBackRefs{\CurrentBib}

\bibitem [\protect \citeauthoryear {%
Mokken%
}{%
Mokken%
}{%
{\protect \APACyear {1997}}%
}]{%
Mokken1997}
\APACinsertmetastar {%
Mokken1997}%
\begin{APACrefauthors}%
Mokken, R\BPBI J.%
\end{APACrefauthors}%
\unskip\
\newblock
\APACrefYearMonthDay{1997}{}{}.
\newblock
{\BBOQ}\APACrefatitle {{Nonparametric Models for Dichotomous Responses}}
  {{Nonparametric Models for Dichotomous Responses}}.{\BBCQ}
\newblock
\BIn{} \APACrefbtitle {Handbook of Modern Item Response Theory.} {Handbook of
  modern item response theory.}
\newblock
\begin{APACrefDOI} \doi{10.1007/978-1-4757-2691-6_20} \end{APACrefDOI}
\PrintBackRefs{\CurrentBib}

\bibitem [\protect \citeauthoryear {%
Musil%
, Warner%
, Yobas%
\BCBL {}\ \BBA {} Jones%
}{%
Musil%
\ \protect \BOthers {.}}{%
{\protect \APACyear {2002}}%
}]{%
Musil2002a}
\APACinsertmetastar {%
Musil2002a}%
\begin{APACrefauthors}%
Musil, C\BPBI M.%
, Warner, C\BPBI B.%
, Yobas, P\BPBI K.%
\BCBL {}\ \BBA {} Jones, S\BPBI L.%
\end{APACrefauthors}%
\unskip\
\newblock
\APACrefYearMonthDay{2002}{}{}.
\newblock
{\BBOQ}\APACrefatitle {{A comparison of imputation techniques for handling
  Missing data}} {{A comparison of imputation techniques for handling Missing
  data}}.{\BBCQ}
\newblock
\APACjournalVolNumPages{Western Journal of Nursing Research}{}{}{}.
\newblock
\begin{APACrefDOI} \doi{10.1177/019394502762477004} \end{APACrefDOI}
\PrintBackRefs{\CurrentBib}

\bibitem [\protect \citeauthoryear {%
Myrtveit%
, Stensrud%
\BCBL {}\ \BBA {} Olsson%
}{%
Myrtveit%
\ \protect \BOthers {.}}{%
{\protect \APACyear {2001}}%
}]{%
Myrtveit2001}
\APACinsertmetastar {%
Myrtveit2001}%
\begin{APACrefauthors}%
Myrtveit, I.%
, Stensrud, E.%
\BCBL {}\ \BBA {} Olsson, U\BPBI H.%
\end{APACrefauthors}%
\unskip\
\newblock
\APACrefYearMonthDay{2001}{}{}.
\newblock
{\BBOQ}\APACrefatitle {{Analyzing data sets with missing data: An empirical
  evaluation of imputation methods and likelihood-based methods}} {{Analyzing
  data sets with missing data: An empirical evaluation of imputation methods
  and likelihood-based methods}}.{\BBCQ}
\newblock
\APACjournalVolNumPages{IEEE Transactions on Software Engineering}{}{}{}.
\newblock
\begin{APACrefDOI} \doi{10.1109/32.965340} \end{APACrefDOI}
\PrintBackRefs{\CurrentBib}

\bibitem [\protect \citeauthoryear {%
Oshiro%
, Perez%
\BCBL {}\ \BBA {} Baranauskas%
}{%
Oshiro%
\ \protect \BOthers {.}}{%
{\protect \APACyear {2012}}%
}]{%
Oshiro2012}
\APACinsertmetastar {%
Oshiro2012}%
\begin{APACrefauthors}%
Oshiro, T\BPBI M.%
, Perez, P\BPBI S.%
\BCBL {}\ \BBA {} Baranauskas, J\BPBI A.%
\end{APACrefauthors}%
\unskip\
\newblock
\APACrefYearMonthDay{2012}{}{}.
\newblock
{\BBOQ}\APACrefatitle {{How many trees in a random forest?}} {{How many trees
  in a random forest?}}{\BBCQ}
\newblock
\BIn{} \APACrefbtitle {Lecture Notes in Computer Science (including subseries
  Lecture Notes in Artificial Intelligence and Lecture Notes in
  Bioinformatics).} {Lecture notes in computer science (including subseries
  lecture notes in artificial intelligence and lecture notes in
  bioinformatics).}
\newblock
\begin{APACrefDOI} \doi{10.1007/978-3-642-31537-4_13} \end{APACrefDOI}
\PrintBackRefs{\CurrentBib}

\bibitem [\protect \citeauthoryear {%
Pantanowitz%
\ \BBA {} Marwala%
}{%
Pantanowitz%
\ \BBA {} Marwala%
}{%
{\protect \APACyear {2009}}%
}]{%
Pantanowitz2009}
\APACinsertmetastar {%
Pantanowitz2009}%
\begin{APACrefauthors}%
Pantanowitz, A.%
\BCBT {}\ \BBA {} Marwala, T.%
\end{APACrefauthors}%
\unskip\
\newblock
\APACrefYearMonthDay{2009}{}{}.
\newblock
{\BBOQ}\APACrefatitle {{Evaluating the impact of missing data imputation}}
  {{Evaluating the impact of missing data imputation}}.{\BBCQ}
\newblock
\BIn{} \APACrefbtitle {Lecture Notes in Computer Science (including subseries
  Lecture Notes in Artificial Intelligence and Lecture Notes in
  Bioinformatics).} {Lecture notes in computer science (including subseries
  lecture notes in artificial intelligence and lecture notes in
  bioinformatics).}
\newblock
\begin{APACrefDOI} \doi{10.1007/978-3-642-03348-3_59} \end{APACrefDOI}
\PrintBackRefs{\CurrentBib}

\bibitem [\protect \citeauthoryear {%
Rubin%
}{%
Rubin%
}{%
{\protect \APACyear {1976}}%
}]{%
Rubin1976}
\APACinsertmetastar {%
Rubin1976}%
\begin{APACrefauthors}%
Rubin, D\BPBI B.%
\end{APACrefauthors}%
\unskip\
\newblock
\APACrefYearMonthDay{1976}{}{}.
\newblock
{\BBOQ}\APACrefatitle {{Inference and Missing Data}} {{Inference and Missing
  Data}}.{\BBCQ}
\newblock
\APACjournalVolNumPages{Biometrika}{}{}{}.
\newblock
\begin{APACrefDOI} \doi{10.2307/2335739} \end{APACrefDOI}
\PrintBackRefs{\CurrentBib}

\bibitem [\protect \citeauthoryear {%
Schafer%
}{%
Schafer%
}{%
{\protect \APACyear {1997}}%
}]{%
Schafer1997}
\APACinsertmetastar {%
Schafer1997}%
\begin{APACrefauthors}%
Schafer, J\BPBI L.%
\end{APACrefauthors}%
\unskip\
\newblock
\APACrefYear{1997}.
\newblock
\APACrefbtitle {{Analysis of Incomplete Multivariate Data}} {{Analysis of
  Incomplete Multivariate Data}}.
\newblock
\APACaddressPublisher{}{London: Chapman {\&} Hall}.
\newblock
\begin{APACrefDOI} \doi{10.1201/9781439821862} \end{APACrefDOI}
\PrintBackRefs{\CurrentBib}

\bibitem [\protect \citeauthoryear {%
Schafer%
}{%
Schafer%
}{%
{\protect \APACyear {1999}}%
}]{%
Schafer1999a}
\APACinsertmetastar {%
Schafer1999a}%
\begin{APACrefauthors}%
Schafer, J\BPBI L.%
\end{APACrefauthors}%
\unskip\
\newblock
\APACrefYearMonthDay{1999}{}{}.
\newblock
{\BBOQ}\APACrefatitle {{Multiple imputation: A primer}} {{Multiple imputation:
  A primer}}.{\BBCQ}
\newblock
\APACjournalVolNumPages{Statistical Methods in Medical Research}{8}{1}{3--15}.
\newblock
\begin{APACrefDOI} \doi{10.1191/096228099671525676} \end{APACrefDOI}
\PrintBackRefs{\CurrentBib}

\bibitem [\protect \citeauthoryear {%
Stekhoven%
}{%
Stekhoven%
}{%
{\protect \APACyear {2012}}%
}]{%
Stekhoven2012}
\APACinsertmetastar {%
Stekhoven2012}%
\begin{APACrefauthors}%
Stekhoven, D\BPBI J.%
\end{APACrefauthors}%
\unskip\
\newblock
\APACrefYearMonthDay{2012}{}{}.
\newblock
{\BBOQ}\APACrefatitle {{Package ''Missforest'}} {{Package
  ''Missforest'}}.{\BBCQ}
\newblock
\APACjournalVolNumPages{Bioinformatics}{}{}{}.
\newblock
\begin{APACrefDOI} \doi{10.1093/bioinformatics/btr597} \end{APACrefDOI}
\PrintBackRefs{\CurrentBib}

\bibitem [\protect \citeauthoryear {%
Su%
, Gelman%
, Hill%
\BCBL {}\ \BBA {} Yajima%
}{%
Su%
\ \protect \BOthers {.}}{%
{\protect \APACyear {2011}}%
}]{%
Su2011}
\APACinsertmetastar {%
Su2011}%
\begin{APACrefauthors}%
Su, Y\BPBI S.%
, Gelman, A.%
, Hill, J.%
\BCBL {}\ \BBA {} Yajima, M.%
\end{APACrefauthors}%
\unskip\
\newblock
\APACrefYearMonthDay{2011}{}{}.
\newblock
{\BBOQ}\APACrefatitle {{Multiple imputation with diagnostics (mi) in R: Opening
  windows into the black box}} {{Multiple imputation with diagnostics (mi) in
  R: Opening windows into the black box}}.{\BBCQ}
\newblock
\APACjournalVolNumPages{Journal of Statistical Software}{}{}{}.
\newblock
\begin{APACrefDOI} \doi{10.18637/jss.v045.i02} \end{APACrefDOI}
\PrintBackRefs{\CurrentBib}

\bibitem [\protect \citeauthoryear {%
Tatsuoka%
, Lord%
, Novick%
\BCBL {}\ \BBA {} Birnbaum%
}{%
Tatsuoka%
\ \protect \BOthers {.}}{%
{\protect \APACyear {1971}}%
}]{%
Tatsuoka1971}
\APACinsertmetastar {%
Tatsuoka1971}%
\begin{APACrefauthors}%
Tatsuoka, M\BPBI M.%
, Lord, F\BPBI M.%
, Novick, M\BPBI R.%
\BCBL {}\ \BBA {} Birnbaum, A.%
\end{APACrefauthors}%
\unskip\
\newblock
\APACrefYearMonthDay{1971}{}{}.
\newblock
{\BBOQ}\APACrefatitle {{Statistical Theories of Mental Test Scores.}}
  {{Statistical Theories of Mental Test Scores.}}{\BBCQ}
\newblock
\APACjournalVolNumPages{Journal of the American Statistical Association}{}{}{}.
\newblock
\begin{APACrefDOI} \doi{10.2307/2283550} \end{APACrefDOI}
\PrintBackRefs{\CurrentBib}

\bibitem [\protect \citeauthoryear {%
{Van Buuren}%
}{%
{Van Buuren}%
}{%
{\protect \APACyear {2012}}%
}]{%
VanBuuren2012}
\APACinsertmetastar {%
VanBuuren2012}%
\begin{APACrefauthors}%
{Van Buuren}, S.%
\end{APACrefauthors}%
\unskip\
\newblock
\APACrefYear{2012}.
\newblock
\APACrefbtitle {{Flexible Imputation of Missing Data}} {{Flexible Imputation of
  Missing Data}}\ (\PrintOrdinal{Second}\ \BEd).
\newblock
\APACaddressPublisher{}{Chapman {\&} Hall/CRC}.
\newblock
\begin{APACrefDOI} \doi{10.1201/b11826} \end{APACrefDOI}
\PrintBackRefs{\CurrentBib}

\bibitem [\protect \citeauthoryear {%
{Van Hulse}%
\ \BBA {} Khoshgoftaar%
}{%
{Van Hulse}%
\ \BBA {} Khoshgoftaar%
}{%
{\protect \APACyear {2008}}%
}]{%
VanHulse2008}
\APACinsertmetastar {%
VanHulse2008}%
\begin{APACrefauthors}%
{Van Hulse}, J.%
\BCBT {}\ \BBA {} Khoshgoftaar, T\BPBI M.%
\end{APACrefauthors}%
\unskip\
\newblock
\APACrefYearMonthDay{2008}{}{}.
\newblock
{\BBOQ}\APACrefatitle {{A comprehensive empirical evaluation of missing value
  imputation in noisy software measurement data}} {{A comprehensive empirical
  evaluation of missing value imputation in noisy software measurement
  data}}.{\BBCQ}
\newblock
\APACjournalVolNumPages{Journal of Systems and Software}{}{}{}.
\newblock
\begin{APACrefDOI} \doi{10.1016/j.jss.2007.07.043} \end{APACrefDOI}
\PrintBackRefs{\CurrentBib}

\end{thebibliography}

\appendix

\renewcommand{\thesubsection}{\Alph{subsection}}
\centering
\begin{landscape}
\section*{Appendices}
\subsection{Results based on RMSE}\label{ssec:RMSEResults}
\tiny
\begin{longtable}[htbp]{ccccccccccccccccccccccc}
		\hline
		&            & \multicolumn{7}{c}{5\% Missing Percentage}                               & \multicolumn{7}{c}{15\% Missing Percentage}                              & \multicolumn{7}{c}{25\% Missing Percentage}                               \\ 
		\cline{3-23}
		Variable & Method     & $\mu$   & Median & Q1     & Q2     & Q3     & Q4     & $\sigma$ & $\mu$   & Median & Q1     & Q2     & Q3     & Q4     & $\sigma$ & $\mu$   & Median & Q1     & Q2     & Q3     & Q4     & $\sigma$  \\ 
		\hline \hline
		\endhead
		\\
		\hline \multicolumn{23}{r}{{continued on next page}}
		\endfoot
		\endlastfoot
		X0       & Mean       & 0.9972 & 0.9961 & 0.9735 & 0.9906 & 1.0047 & 1.0234 & 0.0115             & 0.9982 & 0.9960 & 0.9863 & 0.9938 & 1.0015 & 1.0241 & 0.0073             & 1.0000 & 1.0009 & 0.9831 & 0.9950 & 1.0038 & 1.0136 & 0.0065              \\
		& MICE       & 1.1617 & 1.1626 & 1.1130 & 1.1517 & 1.1725 & 1.2033 & 0.0156             & 1.1812 & 1.1814 & 1.1447 & 1.1749 & 1.1879 & 1.2071 & 0.0100             & 1.2034 & 1.2043 & 1.1802 & 1.1976 & 1.2084 & 1.2245 & 0.0076              \\
		& Mi         & 1.4140 & 1.4123 & 1.3757 & 1.4028 & 1.4265 & 1.4544 & 0.0176             & 1.4123 & 1.4123 & 1.3916 & 1.4060 & 1.4200 & 1.4390 & 0.0099             & 1.4141 & 1.4140 & 1.3988 & 1.4072 & 1.4197 & 1.4295 & 0.0076              \\
		& Amelia     & 1.1784 & 1.1780 & 1.1324 & 1.1675 & 1.1895 & 1.2285 & 0.0161             & 1.1941 & 1.1952 & 1.1563 & 1.1872 & 1.2009 & 1.2188 & 0.0103             & 1.2110 & 1.2110 & 1.1886 & 1.2060 & 1.2158 & 1.2378 & 0.0080              \\
		& Hmisc      & 1.1639 & 1.1633 & 1.1220 & 1.1541 & 1.1732 & 1.2151 & 0.0156             & 1.1835 & 1.1828 & 1.1611 & 1.1757 & 1.1902 & 1.2151 & 0.0101             & 1.2053 & 1.2050 & 1.1865 & 1.2000 & 1.2110 & 1.2264 & 0.0077              \\
		& missForest & 0.7597 & 0.7542 & 0.7213 & 0.7464 & 0.7596 & 1.0013 & 0.0411             & 0.7858 & 0.7854 & 0.7684 & 0.7808 & 0.7901 & 0.8036 & 0.0077             & 0.8129 & 0.8131 & 0.7967 & 0.8090 & 0.8162 & 0.9999 & 0.0135              \\
		X1       & Mean       & 0.9985 & 0.9976 & 0.8353 & 0.9430 & 1.0455 & 1.2137 & 0.0788             & 0.9915 & 0.9894 & 0.8822 & 0.9490 & 1.0407 & 1.1306 & 0.0573             & 1.0046 & 0.9977 & 0.8838 & 0.9696 & 1.0366 & 1.1061 & 0.0496              \\
		& MICE       & 0.5413 & 0.5342 & 0.4323 & 0.5130 & 0.5665 & 0.7632 & 0.0455             & 0.5870 & 0.5850 & 0.5060 & 0.5641 & 0.6089 & 0.7217 & 0.0322             & 0.6486 & 0.6471 & 0.5845 & 0.6297 & 0.6675 & 0.7330 & 0.0277              \\
		& Mi         & 1.4107 & 1.4129 & 1.2638 & 1.3593 & 1.4550 & 1.6400 & 0.0706             & 1.3932 & 1.3904 & 1.3258 & 1.3648 & 1.4259 & 1.4943 & 0.0394             & 1.4400 & 1.4469 & 1.3606 & 1.4059 & 1.4718 & 1.5292 & 0.0398              \\
		& Amelia     & 0.5739 & 0.5676 & 0.5239 & 0.5539 & 0.5895 & 0.6778 & 0.0286             & 0.6165 & 0.6149 & 0.5762 & 0.6019 & 0.6308 & 0.6740 & 0.0193             & 0.6745 & 0.6740 & 0.6281 & 0.6607 & 0.6881 & 0.7263 & 0.0185              \\
		& Hmisc      & 0.5389 & 0.5340 & 0.4422 & 0.5086 & 0.5637 & 0.6717 & 0.0420             & 0.5886 & 0.5852 & 0.5198 & 0.5644 & 0.6123 & 0.6815 & 0.0313             & 0.6535 & 0.6524 & 0.5700 & 0.6293 & 0.6753 & 0.7468 & 0.0337              \\
		& missForest & 0.3596 & 0.3309 & 0.2594 & 0.3069 & 0.3569 & 1.1355 & 0.1364             & 0.4477 & 0.4027 & 0.3426 & 0.3741 & 0.4330 & 1.1060 & 0.1592             & 0.4722 & 0.4596 & 0.3996 & 0.4408 & 0.4834 & 0.9984 & 0.0791              \\
		X2       & Mean       & 0.9964 & 0.9954 & 0.8082 & 0.9509 & 1.0477 & 1.1756 & 0.0793             & 0.9979 & 0.9916 & 0.9255 & 0.9685 & 1.0263 & 1.1026 & 0.0409             & 0.9958 & 0.9910 & 0.9250 & 0.9722 & 1.0185 & 1.0876 & 0.0400              \\
		& MICE       & 0.7471 & 0.7435 & 0.6594 & 0.7232 & 0.7703 & 0.8519 & 0.0360             & 0.7769 & 0.7726 & 0.7208 & 0.7577 & 0.7932 & 0.8577 & 0.0270             & 0.8150 & 0.8161 & 0.7553 & 0.7984 & 0.8315 & 0.8743 & 0.0236              \\
		& Mi         & 1.4014 & 1.3986 & 1.2577 & 1.3487 & 1.4314 & 1.6650 & 0.0745             & 1.3990 & 1.3952 & 1.3359 & 1.3816 & 1.4139 & 1.4599 & 0.0271             & 1.3924 & 1.3873 & 1.3329 & 1.3711 & 1.4138 & 1.4673 & 0.0324              \\
		& Amelia     & 0.7528 & 0.7503 & 0.6933 & 0.7352 & 0.7666 & 0.8310 & 0.0232             & 0.7780 & 0.7762 & 0.7318 & 0.7653 & 0.7888 & 0.8229 & 0.0184             & 0.8133 & 0.8121 & 0.7609 & 0.8000 & 0.8260 & 0.8608 & 0.0178              \\
		& Hmisc      & 0.7505 & 0.7500 & 0.6643 & 0.7245 & 0.7745 & 0.8685 & 0.0369             & 0.7779 & 0.7766 & 0.7118 & 0.7601 & 0.7937 & 0.8416 & 0.0261             & 0.8151 & 0.8158 & 0.7500 & 0.7949 & 0.8324 & 0.9054 & 0.0256              \\
		& missForest & 0.5675 & 0.5455 & 0.4938 & 0.5249 & 0.5776 & 1.0808 & 0.0915             & 0.5775 & 0.5695 & 0.5307 & 0.5540 & 0.5933 & 0.9715 & 0.0430             & 0.6047 & 0.6000 & 0.5621 & 0.5885 & 0.6217 & 0.6520 & 0.0220              \\
		X3       & Mean       & 1.0174 & 0.9971 & 0.8194 & 0.9295 & 1.0606 & 1.4417 & 0.1333             & 0.9864 & 0.9856 & 0.8627 & 0.9429 & 1.0243 & 1.1436 & 0.0676             & 0.9894 & 0.9912 & 0.8446 & 0.9556 & 1.0243 & 1.1043 & 0.0583              \\
		& MICE       & 0.4275 & 0.4224 & 0.3251 & 0.4010 & 0.4539 & 0.5921 & 0.0428             & 0.4940 & 0.4889 & 0.4362 & 0.4715 & 0.5111 & 0.6166 & 0.0316             & 0.5578 & 0.5553 & 0.4918 & 0.5423 & 0.5749 & 0.6257 & 0.0272              \\
		& Mi         & 1.4310 & 1.4060 & 1.3174 & 1.3537 & 1.4730 & 1.8115 & 0.1011             & 1.4003 & 1.3998 & 1.3057 & 1.3674 & 1.4315 & 1.5146 & 0.0496             & 1.4073 & 1.4096 & 1.3047 & 1.3838 & 1.4308 & 1.4910 & 0.0403              \\
		& Amelia     & 0.5136 & 0.5123 & 0.4474 & 0.4967 & 0.5305 & 0.5803 & 0.0237             & 0.5714 & 0.5705 & 0.5211 & 0.5554 & 0.5867 & 0.6284 & 0.0218             & 0.6287 & 0.6271 & 0.5806 & 0.6173 & 0.6422 & 0.6718 & 0.0185              \\
		& Hmisc      & 0.4579 & 0.4349 & 0.3299 & 0.4131 & 0.4674 & 1.0809 & 0.1042             & 0.5070 & 0.5047 & 0.4367 & 0.4849 & 0.5262 & 0.6175 & 0.0324             & 0.5748 & 0.5717 & 0.5170 & 0.5518 & 0.5914 & 0.6641 & 0.0309              \\
		& missForest & 0.3387 & 0.3142 & 0.2151 & 0.2847 & 0.3473 & 1.3226 & 0.1306             & 0.4099 & 0.3720 & 0.3008 & 0.3503 & 0.3937 & 1.0732 & 0.1470             & 0.4433 & 0.4347 & 0.3502 & 0.4133 & 0.4553 & 1.0945 & 0.0888              \\
		X4       & Mean       & 1.0067 & 1.0023 & 0.9467 & 0.9791 & 1.0302 & 1.1145 & 0.0365             & 0.9966 & 0.9954 & 0.9661 & 0.9838 & 1.0090 & 1.0380 & 0.0183             & 1.0006 & 0.9992 & 0.9715 & 0.9884 & 1.0084 & 1.0425 & 0.0158              \\
		& MICE       & 1.1240 & 1.1193 & 1.0614 & 1.1021 & 1.1392 & 1.2487 & 0.0341             & 1.1323 & 1.1328 & 1.0807 & 1.1184 & 1.1438 & 1.2198 & 0.0206             & 1.1566 & 1.1567 & 1.1208 & 1.1455 & 1.1655 & 1.2146 & 0.0158              \\
		& Mi         & 1.4159 & 1.4106 & 1.3615 & 1.3970 & 1.4335 & 1.5046 & 0.0274             & 1.4083 & 1.4059 & 1.3686 & 1.3972 & 1.4197 & 1.4429 & 0.0153             & 1.4120 & 1.4119 & 1.3806 & 1.4032 & 1.4215 & 1.4416 & 0.0118              \\
		& Amelia     & 1.1495 & 1.1464 & 1.0849 & 1.1297 & 1.1650 & 1.2546 & 0.0285             & 1.1566 & 1.1570 & 1.1194 & 1.1456 & 1.1681 & 1.1939 & 0.0162             & 1.1757 & 1.1747 & 1.1397 & 1.1653 & 1.1837 & 1.2269 & 0.0149              \\
		& Hmisc      & 1.1232 & 1.1184 & 1.0516 & 1.1000 & 1.1385 & 1.2479 & 0.0350             & 1.1353 & 1.1346 & 1.0877 & 1.1204 & 1.1481 & 1.2172 & 0.0209             & 1.1584 & 1.1582 & 1.1164 & 1.1466 & 1.1686 & 1.2002 & 0.0160              \\
		& missForest & 0.7635 & 0.7525 & 0.6931 & 0.7392 & 0.7784 & 1.0583 & 0.0499             & 0.7659 & 0.7623 & 0.7277 & 0.7526 & 0.7805 & 0.9975 & 0.0233             & 0.7969 & 0.7894 & 0.7590 & 0.7800 & 0.8018 & 1.0414 & 0.0401              \\
		X5       & Mean       & 1.0153 & 1.0158 & 0.8371 & 0.9286 & 1.0762 & 1.2897 & 0.1016             & 1.0078 & 1.0060 & 0.9119 & 0.9721 & 1.0489 & 1.1225 & 0.0543             & 1.0062 & 1.0028 & 0.9350 & 0.9865 & 1.0277 & 1.0856 & 0.0334              \\
		& MICE       & 0.4408 & 0.4382 & 0.3798 & 0.4206 & 0.4572 & 0.5284 & 0.0282             & 0.5038 & 0.5010 & 0.4557 & 0.4865 & 0.5206 & 0.5765 & 0.0238             & 0.5628 & 0.5631 & 0.5160 & 0.5479 & 0.5760 & 0.6115 & 0.0196              \\
		& Mi         & 1.4287 & 1.4256 & 1.2555 & 1.3703 & 1.4735 & 1.6410 & 0.0804             & 1.4247 & 1.4230 & 1.3585 & 1.3937 & 1.4536 & 1.5155 & 0.0391             & 1.4182 & 1.4143 & 1.3627 & 1.4029 & 1.4350 & 1.4712 & 0.0242              \\
		& Amelia     & 0.4855 & 0.4829 & 0.4348 & 0.4713 & 0.4986 & 0.5380 & 0.0190             & 0.5424 & 0.5425 & 0.5041 & 0.5323 & 0.5525 & 0.5839 & 0.0165             & 0.5952 & 0.5962 & 0.5603 & 0.5849 & 0.6047 & 0.6342 & 0.0147              \\
		& Hmisc      & 0.4569 & 0.4475 & 0.3725 & 0.4320 & 0.4690 & 0.6915 & 0.0495             & 0.5176 & 0.5116 & 0.4527 & 0.4970 & 0.5312 & 0.6399 & 0.0312             & 0.5730 & 0.5720 & 0.5305 & 0.5605 & 0.5845 & 0.6650 & 0.0209              \\
		& missForest & 0.3544 & 0.3395 & 0.2788 & 0.3229 & 0.3668 & 1.2193 & 0.0950             & 0.4268 & 0.3946 & 0.3579 & 0.3770 & 0.4191 & 1.1225 & 0.1311             & 0.4825 & 0.4479 & 0.4107 & 0.4324 & 0.4640 & 1.0856 & 0.1379              \\
		X6       & Mean       & 0.9892 & 0.9590 & 0.8195 & 0.9187 & 1.0532 & 1.2866 & 0.0975             & 1.0085 & 1.0012 & 0.9085 & 0.9697 & 1.0452 & 1.1764 & 0.0558             & 0.9854 & 0.9768 & 0.9021 & 0.9662 & 1.0148 & 1.0861 & 0.0391              \\
		& MICE       & 0.5050 & 0.5012 & 0.4261 & 0.4806 & 0.5259 & 0.6347 & 0.0361             & 0.5584 & 0.5573 & 0.5013 & 0.5417 & 0.5755 & 0.6453 & 0.0259             & 0.6050 & 0.6035 & 0.5517 & 0.5910 & 0.6171 & 0.6716 & 0.0210              \\
		& Mi         & 1.3959 & 1.3821 & 1.2748 & 1.3375 & 1.4403 & 1.6296 & 0.0745             & 1.3756 & 1.3725 & 1.2868 & 1.3456 & 1.4032 & 1.5098 & 0.0431             & 1.3927 & 1.3917 & 1.3311 & 1.3735 & 1.4128 & 1.4645 & 0.0282              \\
		& Amelia     & 0.5083 & 0.5066 & 0.4380 & 0.4919 & 0.5218 & 0.5932 & 0.0265             & 0.5655 & 0.5650 & 0.5152 & 0.5521 & 0.5777 & 0.6100 & 0.0187             & 0.6144 & 0.6117 & 0.5723 & 0.6033 & 0.6225 & 0.6612 & 0.0172              \\
		& Hmisc      & 0.5197 & 0.5031 & 0.4312 & 0.4805 & 0.5352 & 0.7897 & 0.0692             & 0.5618 & 0.5582 & 0.4821 & 0.5436 & 0.5787 & 0.6839 & 0.0279             & 0.6088 & 0.6058 & 0.5577 & 0.5949 & 0.6241 & 0.6891 & 0.0216              \\
		& missForest & 0.4284 & 0.3932 & 0.3224 & 0.3746 & 0.4408 & 1.0874 & 0.1187             & 0.4480 & 0.4394 & 0.3815 & 0.4266 & 0.4584 & 0.9878 & 0.0499             & 0.4847 & 0.4769 & 0.4327 & 0.4595 & 0.4942 & 1.0156 & 0.0597              \\
		X7       & Mean       & 0.9928 & 0.9839 & 0.7673 & 0.9167 & 1.0594 & 1.1870 & 0.0937             & 0.9998 & 0.9889 & 0.8818 & 0.9524 & 1.0398 & 1.1414 & 0.0634             & 1.0002 & 0.9934 & 0.8868 & 0.9606 & 1.0342 & 1.1326 & 0.0544              \\
		& MICE       & 0.6641 & 0.6600 & 0.5133 & 0.6152 & 0.7024 & 0.9439 & 0.0657             & 0.7131 & 0.7101 & 0.6127 & 0.6818 & 0.7427 & 0.8248 & 0.0433             & 0.7596 & 0.7567 & 0.6674 & 0.7366 & 0.7850 & 0.8622 & 0.0357              \\
		& Mi         & 1.4151 & 1.4132 & 1.2803 & 1.3655 & 1.4545 & 1.5587 & 0.0676             & 1.4304 & 1.4193 & 1.3437 & 1.4000 & 1.4626 & 1.5212 & 0.0445             & 1.4188 & 1.4121 & 1.3292 & 1.3951 & 1.4495 & 1.5271 & 0.0444              \\
		& Amelia     & 0.7309 & 0.7322 & 0.6488 & 0.6992 & 0.7602 & 0.8242 & 0.0379             & 0.7708 & 0.7707 & 0.7130 & 0.7519 & 0.7889 & 0.8300 & 0.0266             & 0.8105 & 0.8096 & 0.7468 & 0.7937 & 0.8306 & 0.8886 & 0.0259              \\
		& Hmisc      & 0.6959 & 0.6923 & 0.5355 & 0.6402 & 0.7462 & 1.0265 & 0.0779             & 0.7469 & 0.7450 & 0.6084 & 0.7184 & 0.7753 & 0.8831 & 0.0474             & 0.7873 & 0.7901 & 0.6574 & 0.7636 & 0.8093 & 0.9128 & 0.0389              \\
		& missForest & 0.4284 & 0.4023 & 0.3286 & 0.3821 & 0.4320 & 1.1307 & 0.1191             & 0.5043 & 0.4589 & 0.3742 & 0.4456 & 0.4960 & 1.1414 & 0.1439             & 0.5184 & 0.5071 & 0.4500 & 0.4928 & 0.5381 & 0.9972 & 0.0588              \\
		X8       & Mean       & 0.9913 & 0.9868 & 0.8116 & 0.9169 & 1.0447 & 1.1847 & 0.0930             & 0.9809 & 0.9725 & 0.8647 & 0.9232 & 1.0234 & 1.1813 & 0.0746             & 1.0013 & 0.9949 & 0.8995 & 0.9600 & 1.0383 & 1.1874 & 0.0594              \\
		& MICE       & 0.4993 & 0.4954 & 0.4206 & 0.4757 & 0.5166 & 0.6440 & 0.0334             & 0.5580 & 0.5534 & 0.4898 & 0.5381 & 0.5757 & 0.6752 & 0.0314             & 0.6141 & 0.6110 & 0.5435 & 0.5943 & 0.6326 & 0.7024 & 0.0291              \\
		& Mi         & 1.4089 & 1.4161 & 1.2495 & 1.3355 & 1.4684 & 1.5783 & 0.0882             & 1.3875 & 1.3776 & 1.3000 & 1.3446 & 1.4163 & 1.5442 & 0.0544             & 1.4204 & 1.4150 & 1.3423 & 1.3864 & 1.4521 & 1.5542 & 0.0441              \\
		& Amelia     & 0.5177 & 0.5153 & 0.4657 & 0.5026 & 0.5301 & 0.5954 & 0.0230             & 0.5793 & 0.5785 & 0.5266 & 0.5659 & 0.5939 & 0.6376 & 0.0218             & 0.6392 & 0.6385 & 0.5870 & 0.6261 & 0.6535 & 0.7039 & 0.0225              \\
		& Hmisc      & 0.5057 & 0.5025 & 0.4244 & 0.4797 & 0.5251 & 0.7181 & 0.0374             & 0.5652 & 0.5603 & 0.4831 & 0.5441 & 0.5794 & 0.6746 & 0.0343             & 0.6229 & 0.6175 & 0.5461 & 0.5989 & 0.6374 & 0.8167 & 0.0375              \\
		& missForest & 0.4190 & 0.4016 & 0.3264 & 0.3803 & 0.4195 & 1.0568 & 0.1129             & 0.4498 & 0.4474 & 0.3711 & 0.4199 & 0.4725 & 0.5642 & 0.0396             & 0.4990 & 0.4915 & 0.4433 & 0.4749 & 0.5164 & 0.9937 & 0.0453              \\
		X9       & Mean       & 0.9971 & 1.0018 & 0.9087 & 0.9717 & 1.0239 & 1.0865 & 0.0388             & 0.9960 & 0.9962 & 0.9553 & 0.9833 & 1.0110 & 1.0342 & 0.0196             & 0.9996 & 1.0003 & 0.9628 & 0.9879 & 1.0095 & 1.0471 & 0.0179              \\
		& MICE       & 0.9699 & 0.9693 & 0.8860 & 0.9468 & 0.9886 & 1.0546 & 0.0302             & 1.0100 & 1.0096 & 0.9492 & 0.9986 & 1.0220 & 1.0557 & 0.0181             & 1.0474 & 1.0480 & 1.0039 & 1.0372 & 1.0582 & 1.0926 & 0.0152              \\
		& Mi         & 1.4108 & 1.4121 & 1.2964 & 1.3766 & 1.4403 & 1.5301 & 0.0486             & 1.3989 & 1.4001 & 1.3568 & 1.3864 & 1.4110 & 1.4402 & 0.0186             & 1.4059 & 1.4055 & 1.3662 & 1.3938 & 1.4194 & 1.4547 & 0.0182              \\
		& Amelia     & 0.9805 & 0.9800 & 0.9271 & 0.9679 & 0.9949 & 1.0534 & 0.0204             & 1.0183 & 1.0182 & 0.9815 & 1.0103 & 1.0266 & 1.0573 & 0.0136             & 1.0552 & 1.0552 & 1.0154 & 1.0465 & 1.0632 & 1.0856 & 0.0125              \\
		& Hmisc      & 0.9721 & 0.9701 & 0.8866 & 0.9530 & 0.9919 & 1.0675 & 0.0286             & 1.0110 & 1.0104 & 0.9500 & 0.9997 & 1.0237 & 1.0556 & 0.0174             & 1.0493 & 1.0497 & 1.0017 & 1.0375 & 1.0610 & 1.0937 & 0.0165              \\
		& missForest & 0.6838 & 0.6815 & 0.6195 & 0.6638 & 0.6995 & 0.9794 & 0.0392             & 0.7159 & 0.7145 & 0.6832 & 0.7051 & 0.7213 & 1.0147 & 0.0274             & 0.7504 & 0.7455 & 0.7087 & 0.7336 & 0.7563 & 1.0229 & 0.0416              \\
		X10      & Mean       & 0.9935 & 0.9992 & 0.9036 & 0.9699 & 1.0213 & 1.0628 & 0.0395             & 1.0016 & 1.0016 & 0.9492 & 0.9865 & 1.0148 & 1.0598 & 0.0235             & 0.9938 & 0.9942 & 0.9539 & 0.9810 & 1.0068 & 1.0366 & 0.0188              \\
		& MICE       & 0.6581 & 0.6560 & 0.5934 & 0.6434 & 0.6737 & 0.7361 & 0.0224             & 0.6951 & 0.6949 & 0.6541 & 0.6871 & 0.7046 & 0.7288 & 0.0145             & 0.7338 & 0.7331 & 0.6999 & 0.7254 & 0.7421 & 0.7691 & 0.0128              \\
		& Mi         & 1.4102 & 1.4139 & 1.2939 & 1.3864 & 1.4345 & 1.5169 & 0.0403             & 1.4080 & 1.4087 & 1.3678 & 1.3960 & 1.4176 & 1.4512 & 0.0168             & 1.4079 & 1.4068 & 1.3786 & 1.3988 & 1.4171 & 1.4373 & 0.0131              \\
		& Amelia     & 0.6804 & 0.6818 & 0.6321 & 0.6678 & 0.6904 & 0.7294 & 0.0165             & 0.7144 & 0.7146 & 0.6867 & 0.7068 & 0.7225 & 0.7417 & 0.0109             & 0.7483 & 0.7484 & 0.7221 & 0.7419 & 0.7540 & 0.7808 & 0.0103              \\
		& Hmisc      & 0.6612 & 0.6621 & 0.6017 & 0.6465 & 0.6745 & 0.7432 & 0.0225             & 0.6980 & 0.6977 & 0.6524 & 0.6891 & 0.7078 & 0.7434 & 0.0147             & 0.7374 & 0.7358 & 0.7056 & 0.7279 & 0.7466 & 0.7792 & 0.0135              \\
		& missForest & 0.4572 & 0.4450 & 0.4024 & 0.4333 & 0.4612 & 1.0017 & 0.0748             & 0.4934 & 0.4911 & 0.4597 & 0.4828 & 0.4992 & 0.9592 & 0.0326             & 0.5401 & 0.5267 & 0.4967 & 0.5168 & 0.5350 & 1.0202 & 0.0832              \\
		X11      & Mean       & 1.0012 & 0.9995 & 0.9414 & 0.9841 & 1.0208 & 1.0571 & 0.0252             & 1.0018 & 1.0004 & 0.9696 & 0.9931 & 1.0106 & 1.0420 & 0.0134             & 1.0011 & 0.9999 & 0.9784 & 0.9886 & 1.0099 & 1.0381 & 0.0145              \\
		& MICE       & 0.4052 & 0.4017 & 0.3559 & 0.3901 & 0.4194 & 0.4698 & 0.0216             & 0.5699 & 0.5689 & 0.5296 & 0.5615 & 0.5791 & 0.6146 & 0.0139             & 0.6776 & 0.6770 & 0.6413 & 0.6694 & 0.6859 & 0.7206 & 0.0122              \\
		& Mi         & 1.4172 & 1.4168 & 1.3442 & 1.3944 & 1.4413 & 1.4939 & 0.0299             & 1.4298 & 1.4282 & 1.3978 & 1.4234 & 1.4397 & 1.4630 & 0.0126             & 1.4145 & 1.4132 & 1.3838 & 1.4054 & 1.4249 & 1.4458 & 0.0142              \\
		& Amelia     & 0.3872 & 0.3864 & 0.3474 & 0.3779 & 0.3972 & 0.4380 & 0.0151             & 0.5340 & 0.5342 & 0.4982 & 0.5284 & 0.5391 & 0.5594 & 0.0096             & 0.6441 & 0.6441 & 0.6209 & 0.6365 & 0.6499 & 0.6880 & 0.0104              \\
		& Hmisc      & 0.3997 & 0.3981 & 0.3649 & 0.3868 & 0.4118 & 0.4493 & 0.0173             & 0.5602 & 0.5594 & 0.5250 & 0.5503 & 0.5699 & 0.6033 & 0.0151             & 0.6671 & 0.6672 & 0.6360 & 0.6564 & 0.6791 & 0.6993 & 0.0142              \\
		& missForest & 0.2979 & 0.2847 & 0.2516 & 0.2728 & 0.2981 & 1.0180 & 0.0924             & 0.4012 & 0.3860 & 0.3555 & 0.3805 & 0.3938 & 1.0169 & 0.0940             & 0.4690 & 0.4569 & 0.4357 & 0.4506 & 0.4631 & 1.0261 & 0.0777              \\
		X12      & Mean       & 1.0022 & 1.0001 & 0.9252 & 0.9691 & 1.0249 & 1.1141 & 0.0403             & 1.0006 & 1.0033 & 0.9375 & 0.9796 & 1.0244 & 1.0643 & 0.0309             & 0.9957 & 0.9950 & 0.9601 & 0.9762 & 1.0136 & 1.0544 & 0.0220              \\
		& MICE       & 1.3083 & 1.3056 & 1.1982 & 1.2775 & 1.3395 & 1.4271 & 0.0448             & 1.3161 & 1.3168 & 1.2297 & 1.2967 & 1.3358 & 1.3823 & 0.0284             & 1.3201 & 1.3208 & 1.2600 & 1.3054 & 1.3350 & 1.3807 & 0.0226              \\
		& Mi         & 1.4218 & 1.4102 & 1.2939 & 1.3844 & 1.4789 & 1.5636 & 0.0601             & 1.4389 & 1.4416 & 1.3828 & 1.4238 & 1.4564 & 1.4860 & 0.0224             & 1.4104 & 1.4084 & 1.3634 & 1.3959 & 1.4252 & 1.4686 & 0.0229              \\
		& Amelia     & 1.3109 & 1.3077 & 1.2270 & 1.2891 & 1.3301 & 1.4122 & 0.0335             & 1.3179 & 1.3191 & 1.2477 & 1.3019 & 1.3336 & 1.3843 & 0.0229             & 1.3227 & 1.3232 & 1.2870 & 1.3092 & 1.3335 & 1.3719 & 0.0179              \\
		& Hmisc      & 1.3046 & 1.2969 & 1.1825 & 1.2722 & 1.3308 & 1.4349 & 0.0447             & 1.3139 & 1.3134 & 1.2275 & 1.2951 & 1.3324 & 1.3841 & 0.0275             & 1.3189 & 1.3182 & 1.2637 & 1.3057 & 1.3317 & 1.3705 & 0.0205              \\
		& missForest & 0.9173 & 0.9112 & 0.8504 & 0.8895 & 0.9394 & 1.0446 & 0.0377             & 0.9276 & 0.9307 & 0.8622 & 0.9062 & 0.9488 & 0.9889 & 0.0287             & 0.9346 & 0.9331 & 0.8948 & 0.9187 & 0.9482 & 0.9959 & 0.0203              \\
		X13      & Mean       & 1.0047 & 1.0039 & 0.9784 & 0.9968 & 1.0126 & 1.0334 & 0.0131             & 0.9995 & 0.9991 & 0.9838 & 0.9937 & 1.0051 & 1.0239 & 0.0089             & 1.0006 & 1.0007 & 0.9858 & 0.9947 & 1.0063 & 1.0190 & 0.0084              \\
		& MICE       & 1.1408 & 1.1412 & 1.0622 & 1.1264 & 1.1580 & 1.1948 & 0.0235             & 1.1530 & 1.1531 & 1.1222 & 1.1442 & 1.1619 & 1.1835 & 0.0119             & 1.1682 & 1.1680 & 1.1385 & 1.1621 & 1.1747 & 1.2005 & 0.0097              \\
		& Mi         & 1.4127 & 1.4124 & 1.3569 & 1.3959 & 1.4355 & 1.4727 & 0.0245             & 1.4034 & 1.4027 & 1.3813 & 1.3979 & 1.4086 & 1.4266 & 0.0098             & 1.4085 & 1.4081 & 1.3881 & 1.4029 & 1.4153 & 1.4285 & 0.0085              \\
		& Amelia     & 1.1631 & 1.1628 & 1.1176 & 1.1509 & 1.1759 & 1.2096 & 0.0180             & 1.1726 & 1.1719 & 1.1428 & 1.1673 & 1.1787 & 1.1976 & 0.0100             & 1.1879 & 1.1874 & 1.1598 & 1.1815 & 1.1942 & 1.2187 & 0.0094              \\
		& Hmisc      & 1.1412 & 1.1418 & 1.0803 & 1.1247 & 1.1561 & 1.2170 & 0.0224             & 1.1541 & 1.1547 & 1.1182 & 1.1472 & 1.1621 & 1.1937 & 0.0120             & 1.1712 & 1.1711 & 1.1378 & 1.1634 & 1.1777 & 1.2018 & 0.0107              \\
		& missForest & 0.7578 & 0.7548 & 0.7170 & 0.7423 & 0.7654 & 1.0181 & 0.0357             & 0.7725 & 0.7729 & 0.7474 & 0.7667 & 0.7780 & 0.7989 & 0.0093             & 0.7933 & 0.7928 & 0.7785 & 0.7864 & 0.7973 & 1.0118 & 0.0159              \\
		X14      & Mean       & 1.0058 & 1.0037 & 0.9618 & 0.9883 & 1.0208 & 1.0886 & 0.0257             & 0.9976 & 0.9996 & 0.9607 & 0.9875 & 1.0071 & 1.0334 & 0.0159             & 0.9993 & 0.9974 & 0.9789 & 0.9903 & 1.0093 & 1.0262 & 0.0123              \\
		& MICE       & 0.4067 & 0.4045 & 0.3611 & 0.3920 & 0.4195 & 0.4722 & 0.0207             & 0.5662 & 0.5672 & 0.5331 & 0.5564 & 0.5739 & 0.6082 & 0.0134             & 0.6758 & 0.6754 & 0.6365 & 0.6674 & 0.6836 & 0.7087 & 0.0125              \\
		& Mi         & 1.4144 & 1.4177 & 1.3280 & 1.3981 & 1.4317 & 1.4838 & 0.0287             & 1.4112 & 1.4109 & 1.3826 & 1.4011 & 1.4195 & 1.4455 & 0.0140             & 1.4160 & 1.4163 & 1.3892 & 1.4074 & 1.4246 & 1.4376 & 0.0113              \\
		& Amelia     & 0.3863 & 0.3850 & 0.3492 & 0.3758 & 0.3965 & 0.4286 & 0.0151             & 0.5298 & 0.5293 & 0.4975 & 0.5242 & 0.5356 & 0.5622 & 0.0098             & 0.6422 & 0.6421 & 0.6110 & 0.6351 & 0.6485 & 0.6819 & 0.0106              \\
		& Hmisc      & 0.4052 & 0.4007 & 0.3591 & 0.3898 & 0.4138 & 0.5268 & 0.0260             & 0.5575 & 0.5563 & 0.5185 & 0.5479 & 0.5666 & 0.6064 & 0.0147             & 0.6650 & 0.6664 & 0.6330 & 0.6554 & 0.6752 & 0.6947 & 0.0139              \\
		& missForest & 0.3118 & 0.2903 & 0.2615 & 0.2783 & 0.3004 & 1.0208 & 0.1191             & 0.3886 & 0.3843 & 0.3489 & 0.3786 & 0.3933 & 0.9986 & 0.0411             & 0.4603 & 0.4591 & 0.4359 & 0.4539 & 0.4675 & 0.4879 & 0.0095              \\
		X15      & Mean       & 0.9979 & 0.9978 & 0.9238 & 0.9772 & 1.0218 & 1.1070 & 0.0343             & 0.9999 & 1.0017 & 0.9588 & 0.9857 & 1.0129 & 1.0436 & 0.0190             & 0.9946 & 0.9942 & 0.9648 & 0.9803 & 1.0016 & 1.0308 & 0.0166              \\
		& MICE       & 1.1375 & 1.1349 & 1.0275 & 1.1174 & 1.1560 & 1.2488 & 0.0330             & 1.1641 & 1.1624 & 1.0974 & 1.1492 & 1.1767 & 1.2689 & 0.0232             & 1.1815 & 1.1803 & 1.1386 & 1.1707 & 1.1916 & 1.2401 & 0.0158              \\
		& Mi         & 1.4055 & 1.4043 & 1.3098 & 1.3745 & 1.4382 & 1.4835 & 0.0403             & 1.4115 & 1.4126 & 1.3750 & 1.3995 & 1.4236 & 1.4537 & 0.0166             & 1.4190 & 1.4192 & 1.3807 & 1.4039 & 1.4305 & 1.4670 & 0.0183              \\
		& Amelia     & 1.1481 & 1.1457 & 1.0823 & 1.1300 & 1.1658 & 1.2421 & 0.0264             & 1.1668 & 1.1677 & 1.1224 & 1.1564 & 1.1765 & 1.2169 & 0.0158             & 1.1835 & 1.1831 & 1.1460 & 1.1740 & 1.1926 & 1.2290 & 0.0143              \\
		& Hmisc      & 1.1374 & 1.1331 & 1.0413 & 1.1113 & 1.1582 & 1.2983 & 0.0370             & 1.1634 & 1.1624 & 1.1177 & 1.1505 & 1.1771 & 1.2178 & 0.0200             & 1.1811 & 1.1797 & 1.1364 & 1.1695 & 1.1916 & 1.2316 & 0.0167              \\
		& missForest & 0.7759 & 0.7646 & 0.7078 & 0.7422 & 0.7934 & 1.0241 & 0.0559             & 0.7970 & 0.7942 & 0.7590 & 0.7864 & 0.8052 & 0.9906 & 0.0221             & 0.8145 & 0.8118 & 0.7839 & 0.8042 & 0.8238 & 0.8562 & 0.0149              \\
		X16      & Mean       & 1.0144 & 0.9966 & 0.9162 & 0.9817 & 1.0236 & 1.6872 & 0.1020             & 1.0121 & 1.0009 & 0.9482 & 0.9875 & 1.0119 & 1.3032 & 0.0618             & 0.9946 & 0.9953 & 0.9227 & 0.9855 & 1.0094 & 1.0483 & 0.0246              \\
		& MICE       & 1.1489 & 1.1312 & 1.0378 & 1.1093 & 1.1534 & 1.8268 & 0.1045             & 1.1591 & 1.1490 & 1.0765 & 1.1335 & 1.1604 & 1.4828 & 0.0568             & 1.1669 & 1.1680 & 1.0833 & 1.1565 & 1.1789 & 1.2753 & 0.0267              \\
		& Mi         & 1.4224 & 1.4127 & 1.3360 & 1.3818 & 1.4314 & 2.0013 & 0.0912             & 1.4183 & 1.4086 & 1.3634 & 1.4006 & 1.4226 & 1.6385 & 0.0466             & 1.4031 & 1.4016 & 1.3613 & 1.3921 & 1.4178 & 1.4415 & 0.0186              \\
		& Amelia     & 1.1542 & 1.1408 & 1.0561 & 1.1251 & 1.1548 & 1.7686 & 0.0896             & 1.1674 & 1.1549 & 1.1128 & 1.1461 & 1.1685 & 1.4107 & 0.0515             & 1.1725 & 1.1727 & 1.0877 & 1.1643 & 1.1831 & 1.2220 & 0.0192              \\
		& Hmisc      & 1.1450 & 1.1306 & 1.0227 & 1.1131 & 1.1512 & 1.7348 & 0.0929             & 1.1586 & 1.1490 & 1.0737 & 1.1343 & 1.1664 & 1.4089 & 0.0523             & 1.1649 & 1.1672 & 1.0652 & 1.1539 & 1.1772 & 1.2523 & 0.0251              \\
		& missForest & 0.7819 & 0.7532 & 0.7051 & 0.7397 & 0.7783 & 1.5344 & 0.1201             & 0.7993 & 0.7822 & 0.7351 & 0.7731 & 0.7991 & 1.1224 & 0.0694             & 0.8012 & 0.8028 & 0.7448 & 0.7923 & 0.8125 & 0.8544 & 0.0202              \\
		X17      & Mean       & 1.0000 & 1.0037 & 0.9592 & 0.9860 & 1.0179 & 1.0418 & 0.0218             & 0.9972 & 0.9950 & 0.9671 & 0.9886 & 1.0050 & 1.0294 & 0.0124             & 0.9977 & 0.9961 & 0.9696 & 0.9886 & 1.0058 & 1.0237 & 0.0121              \\
		& MICE       & 0.8675 & 0.8683 & 0.8232 & 0.8572 & 0.8789 & 0.9197 & 0.0169             & 0.9038 & 0.9028 & 0.8771 & 0.8955 & 0.9118 & 0.9309 & 0.0113             & 0.9445 & 0.9449 & 0.9140 & 0.9375 & 0.9507 & 0.9711 & 0.0102              \\
		& Mi         & 1.4110 & 1.4147 & 1.3578 & 1.3943 & 1.4294 & 1.4729 & 0.0242             & 1.4252 & 1.4247 & 1.3926 & 1.4180 & 1.4334 & 1.4516 & 0.0114             & 1.4091 & 1.4078 & 1.3754 & 1.3995 & 1.4178 & 1.4423 & 0.0132              \\
		& Amelia     & 0.8679 & 0.8683 & 0.8253 & 0.8550 & 0.8815 & 0.9094 & 0.0163             & 0.9032 & 0.9025 & 0.8827 & 0.8967 & 0.9091 & 0.9320 & 0.0094             & 0.9428 & 0.9422 & 0.9165 & 0.9350 & 0.9481 & 0.9850 & 0.0111              \\
		& Hmisc      & 0.8683 & 0.8688 & 0.8181 & 0.8570 & 0.8801 & 0.9266 & 0.0176             & 0.9049 & 0.9039 & 0.8780 & 0.8954 & 0.9130 & 0.9403 & 0.0122             & 0.9465 & 0.9461 & 0.9160 & 0.9408 & 0.9524 & 0.9782 & 0.0105              \\
		& missForest & 0.6102 & 0.6055 & 0.5682 & 0.5888 & 0.6142 & 1.0133 & 0.0554             & 0.6342 & 0.6342 & 0.6187 & 0.6269 & 0.6410 & 0.6548 & 0.0091             & 0.6668 & 0.6668 & 0.6406 & 0.6596 & 0.6725 & 0.6923 & 0.0099              \\
		X18      & Mean       & 1.0026 & 0.9965 & 0.9580 & 0.9829 & 1.0192 & 1.0682 & 0.0254             & 0.9976 & 0.9987 & 0.9713 & 0.9872 & 1.0074 & 1.0218 & 0.0133             & 1.0016 & 1.0011 & 0.9873 & 0.9940 & 1.0074 & 1.0292 & 0.0094              \\
		& MICE       & 0.8377 & 0.8371 & 0.7907 & 0.8240 & 0.8506 & 0.8917 & 0.0211             & 0.8752 & 0.8750 & 0.8464 & 0.8674 & 0.8833 & 0.9044 & 0.0119             & 0.9214 & 0.9209 & 0.8945 & 0.9153 & 0.9277 & 0.9456 & 0.0096              \\
		& Mi         & 1.4165 & 1.4182 & 1.3732 & 1.3967 & 1.4323 & 1.4740 & 0.0235             & 1.4130 & 1.4127 & 1.3817 & 1.4049 & 1.4225 & 1.4425 & 0.0124             & 1.4134 & 1.4123 & 1.3902 & 1.4086 & 1.4196 & 1.4392 & 0.0103              \\
		& Amelia     & 0.8387 & 0.8395 & 0.7924 & 0.8263 & 0.8503 & 0.8971 & 0.0175             & 0.8765 & 0.8761 & 0.8499 & 0.8704 & 0.8834 & 0.9050 & 0.0099             & 0.9217 & 0.9210 & 0.9005 & 0.9154 & 0.9275 & 0.9489 & 0.0088              \\
		& Hmisc      & 0.8403 & 0.8400 & 0.7863 & 0.8258 & 0.8554 & 0.9028 & 0.0217             & 0.8770 & 0.8768 & 0.8443 & 0.8686 & 0.8849 & 0.9068 & 0.0119             & 0.9226 & 0.9224 & 0.8953 & 0.9156 & 0.9291 & 0.9514 & 0.0094              \\
		& missForest & 0.6033 & 0.5915 & 0.5497 & 0.5766 & 0.6073 & 1.0047 & 0.0683             & 0.6442 & 0.6235 & 0.5987 & 0.6153 & 0.6310 & 1.0138 & 0.0861             & 0.6657 & 0.6585 & 0.6404 & 0.6539 & 0.6643 & 1.0084 & 0.0488              \\
		X19      & Mean       & 1.0015 & 1.0012 & 0.9610 & 0.9881 & 1.0086 & 1.0460 & 0.0185             & 0.9997 & 0.9992 & 0.9743 & 0.9922 & 1.0078 & 1.0263 & 0.0113             & 1.0017 & 1.0013 & 0.9878 & 0.9970 & 1.0055 & 1.0209 & 0.0082              \\
		& MICE       & 1.3533 & 1.3530 & 1.2819 & 1.3368 & 1.3696 & 1.4101 & 0.0237             & 1.3558 & 1.3560 & 1.3193 & 1.3461 & 1.3661 & 1.3929 & 0.0137             & 1.3612 & 1.3614 & 1.3370 & 1.3546 & 1.3678 & 1.3912 & 0.0109              \\
		& Mi         & 1.4142 & 1.4132 & 1.3448 & 1.3988 & 1.4304 & 1.4795 & 0.0253             & 1.4142 & 1.4129 & 1.3839 & 1.4069 & 1.4236 & 1.4432 & 0.0129             & 1.4121 & 1.4102 & 1.3911 & 1.4056 & 1.4186 & 1.4461 & 0.0099              \\
		& Amelia     & 1.3564 & 1.3555 & 1.2798 & 1.3421 & 1.3717 & 1.4118 & 0.0224             & 1.3558 & 1.3558 & 1.3074 & 1.3474 & 1.3640 & 1.4019 & 0.0130             & 1.3605 & 1.3606 & 1.3331 & 1.3538 & 1.3670 & 1.3887 & 0.0108              \\
		& Hmisc      & 1.3535 & 1.3529 & 1.2801 & 1.3373 & 1.3708 & 1.4173 & 0.0244             & 1.3550 & 1.3548 & 1.3168 & 1.3453 & 1.3643 & 1.3947 & 0.0143             & 1.3611 & 1.3608 & 1.3324 & 1.3543 & 1.3682 & 1.3951 & 0.0103              \\
		& missForest & 0.9583 & 0.9563 & 0.9172 & 0.9435 & 0.9702 & 1.0460 & 0.0209             & 0.9601 & 0.9589 & 0.9373 & 0.9514 & 0.9680 & 0.9880 & 0.0111             & 0.9675 & 0.9672 & 0.9507 & 0.9618 & 0.9710 & 1.0085 & 0.0088              \\
		X20      & Mean       & 0.9987 & 0.9992 & 0.9643 & 0.9904 & 1.0061 & 1.0303 & 0.0138             & 0.9994 & 0.9995 & 0.9822 & 0.9916 & 1.0055 & 1.0232 & 0.0095             & 1.0002 & 1.0006 & 0.9864 & 0.9955 & 1.0037 & 1.0224 & 0.0071              \\
		& MICE       & 0.8164 & 0.8171 & 0.7766 & 0.8071 & 0.8251 & 0.8532 & 0.0132             & 0.8398 & 0.8400 & 0.8204 & 0.8344 & 0.8445 & 0.8608 & 0.0080             & 0.8714 & 0.8713 & 0.8563 & 0.8670 & 0.8759 & 0.8893 & 0.0062              \\
		& Mi         & 1.4108 & 1.4096 & 1.3697 & 1.4006 & 1.4190 & 1.4762 & 0.0176             & 1.4113 & 1.4108 & 1.3890 & 1.4061 & 1.4174 & 1.4376 & 0.0095             & 1.4153 & 1.4151 & 1.3997 & 1.4094 & 1.4211 & 1.4344 & 0.0072              \\
		& Amelia     & 0.8507 & 0.8511 & 0.8153 & 0.8410 & 0.8605 & 0.8812 & 0.0135             & 0.8698 & 0.8705 & 0.8459 & 0.8656 & 0.8747 & 0.8949 & 0.0080             & 0.8968 & 0.8959 & 0.8789 & 0.8918 & 0.9016 & 0.9157 & 0.0067              \\
		& Hmisc      & 0.8190 & 0.8193 & 0.7823 & 0.8105 & 0.8279 & 0.8566 & 0.0126             & 0.8427 & 0.8428 & 0.8183 & 0.8371 & 0.8477 & 0.8637 & 0.0079             & 0.8740 & 0.8739 & 0.8548 & 0.8697 & 0.8783 & 0.8954 & 0.0066              \\
		& missForest & 0.5262 & 0.5109 & 0.4911 & 0.5041 & 0.5172 & 1.0257 & 0.0867             & 0.5399 & 0.5396 & 0.5285 & 0.5357 & 0.5435 & 0.5533 & 0.0059             & 0.5734 & 0.5721 & 0.5595 & 0.5691 & 0.5743 & 0.9962 & 0.0271              \\
		\hline
	\end{longtable}	

\end{landscape}


\centering
\subsection{Box Plots based on RMSE}\label{ssec:RMSEBoxPlots}
  \begin{figure}[h]
		\raggedleft
		\addtocounter{subfigure}{-1}
		\subfigure{ \frame{\includegraphics[width=0.33\textwidth ]{Images/Legends}}}
		\begin{center}
		\subfigure[Performance based on RMSE: X0]{%
			\label{fig:first}
			\frame{\includegraphics[width=0.33\textwidth]{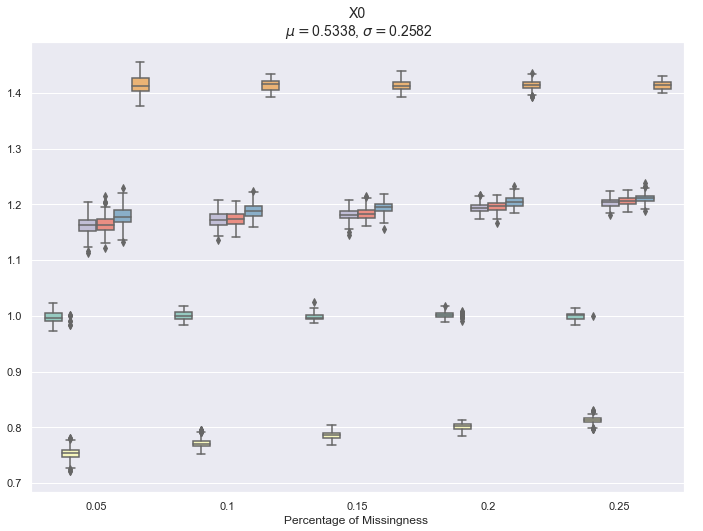}}
		}%
		\subfigure[Performance based on RMSE: X1]{%
			\label{fig:second}
			\frame{\includegraphics[width=0.33\textwidth]{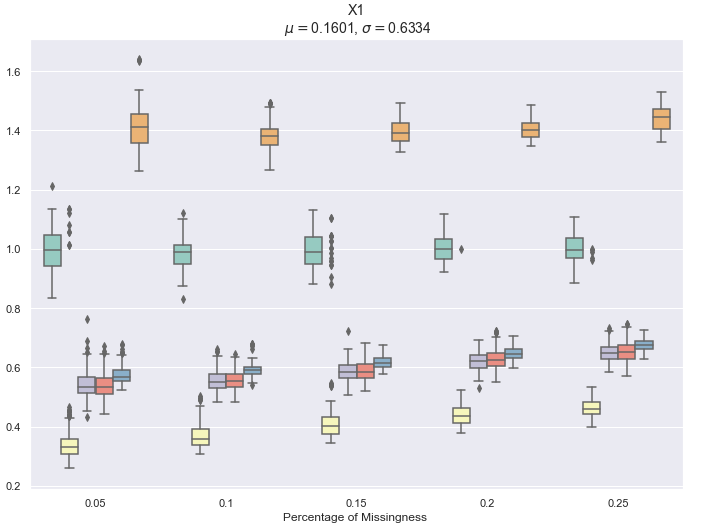}}
		}%
		\subfigure[Performance based on RMSE: X2]{%
			\label{fig:fourth}
			\frame{\includegraphics[width=0.33\textwidth]{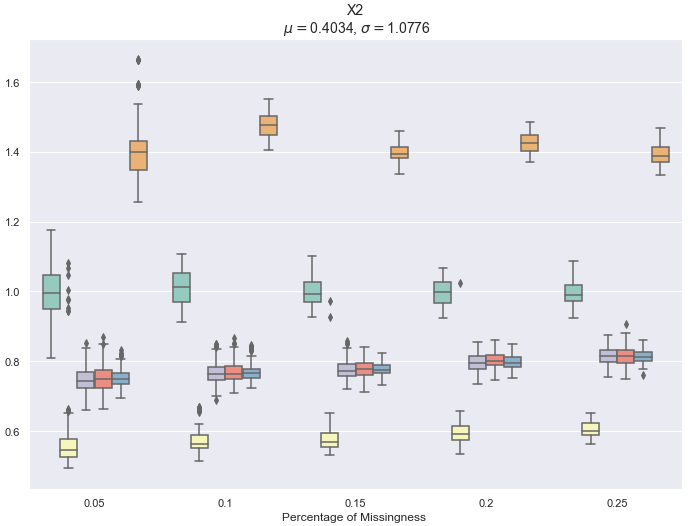}}
		}%
		\\

		\subfigure[Performance based on RMSE: X3]{%
			\label{fig:first}
			\frame{\includegraphics[width=0.33\textwidth]{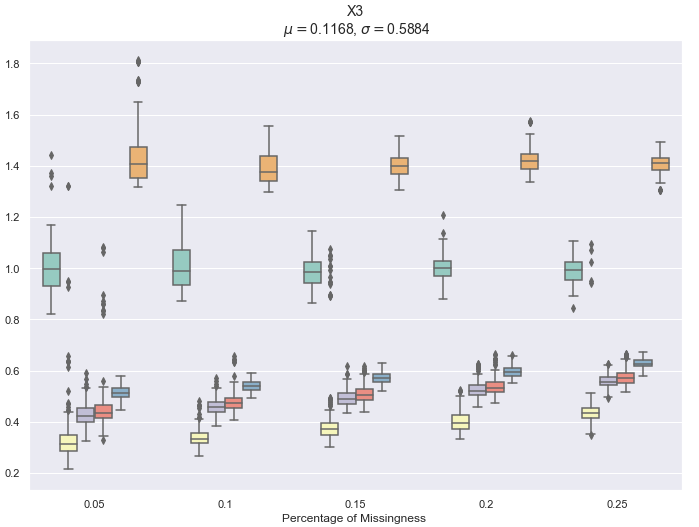}}
		}%
		\subfigure[Performance based on RMSE: X4]{%
			\label{fig:second}
			\frame{\includegraphics[width=0.33\textwidth]{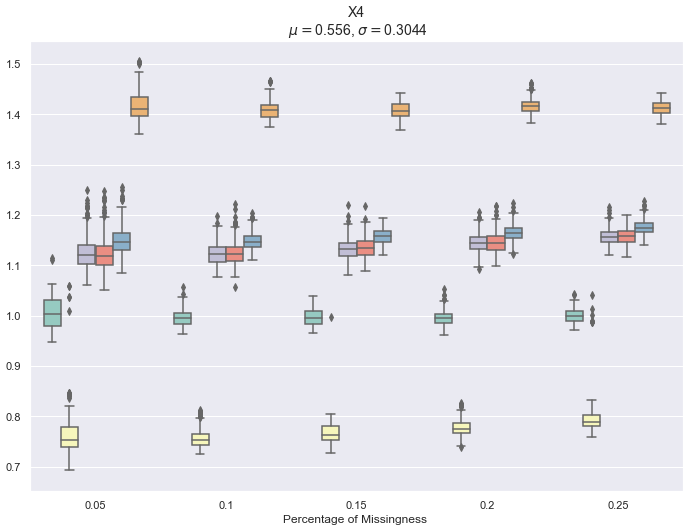}}
		}%
		\subfigure[Performance based on RMSE: X5]{%
			\label{fig:fourth}
			\frame{\includegraphics[width=0.33\textwidth]{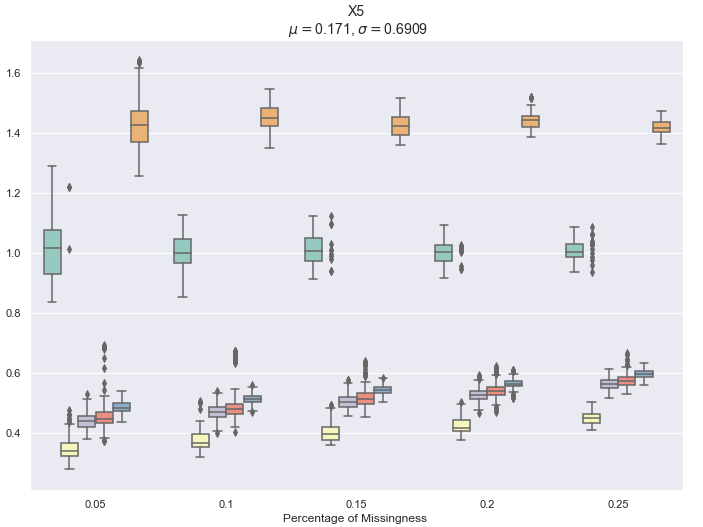}}
		}%
		\\
				\subfigure[Performance based on RMSE: X6]{%
			\label{fig:first}
			\frame{\includegraphics[width=0.33\textwidth]{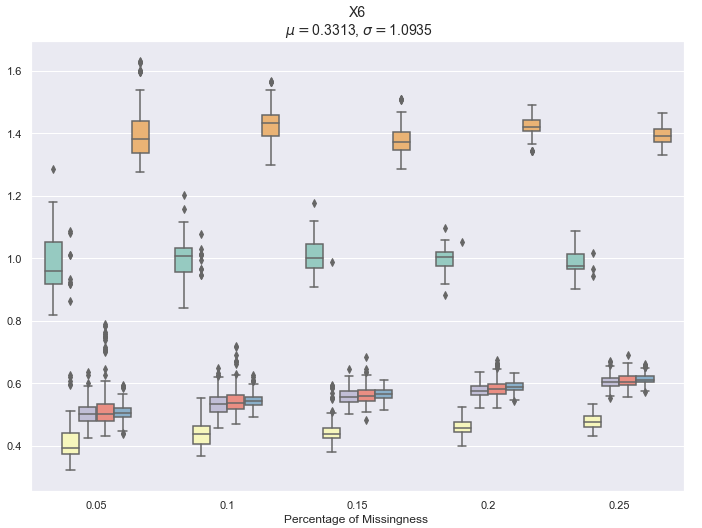}}
		}%
		\subfigure[Performance based on RMSE: X8]{%
			\label{fig:second}
			\frame{\includegraphics[width=0.33\textwidth]{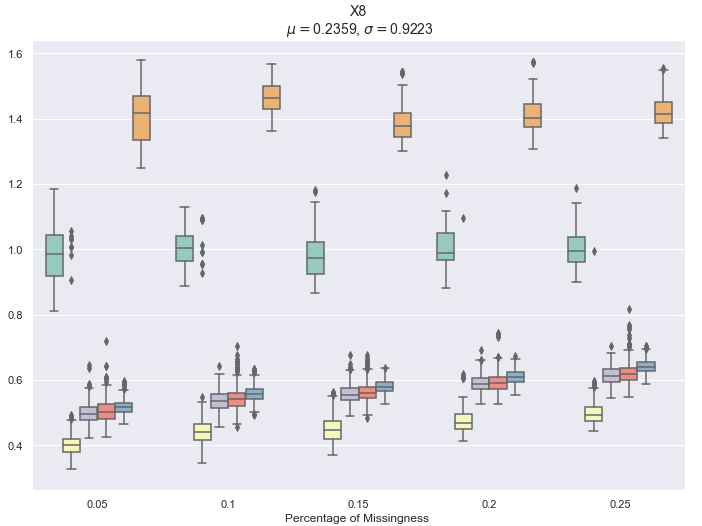}}
		}%
		\subfigure[Performance based on RMSE: X10]{%
			\label{fig:fourth}
			\frame{\includegraphics[width=0.33\textwidth]{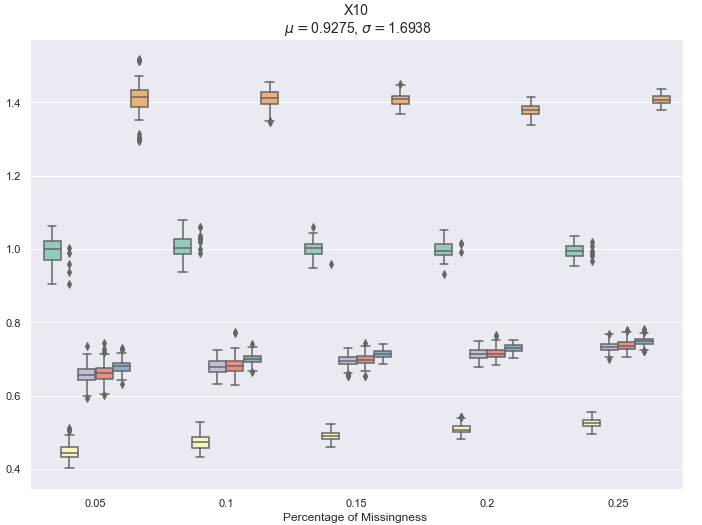}}
		}%

        \end{center}
			\label{fig:ThreeBoxPlots}
\end{figure}

  \newpage
  \begin{figure*}[!]
		\raggedleft
		\addtocounter{subfigure}{-1}
		\subfigure{ \frame{\includegraphics[width=0.33\textwidth ]{Images/Legends}}}
		\begin{center}
	    \subfigure[Performance based on RMSE: X11]{%
			\label{fig:first}
			\frame{\includegraphics[width=0.33\textwidth]{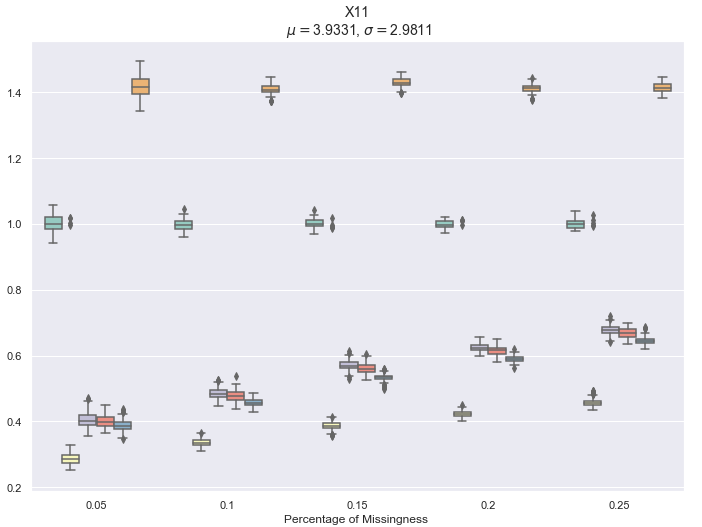}}
		}%
		\subfigure[Performance based on RMSE: X13]{%
			\label{fig:second}
			\frame{\includegraphics[width=0.33\textwidth]{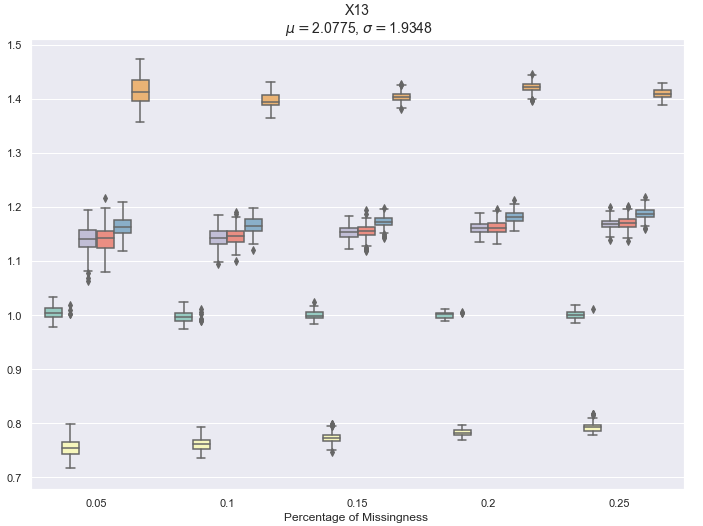}}
		}%
		\subfigure[Performance based on RMSE: X14]{%
			\label{fig:fourth}
			\frame{\includegraphics[width=0.33\textwidth]{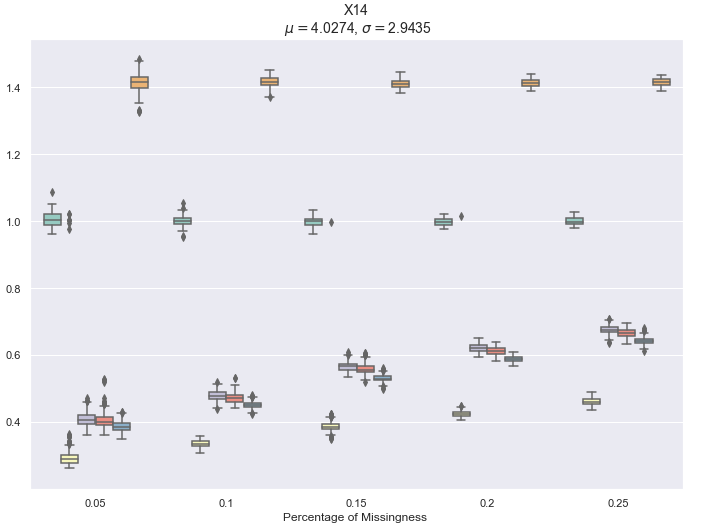}}
		}%
	\\
		\subfigure[Performance based on RMSE: X15]{%
			\label{fig:first}
			\frame{\includegraphics[width=0.33\textwidth]{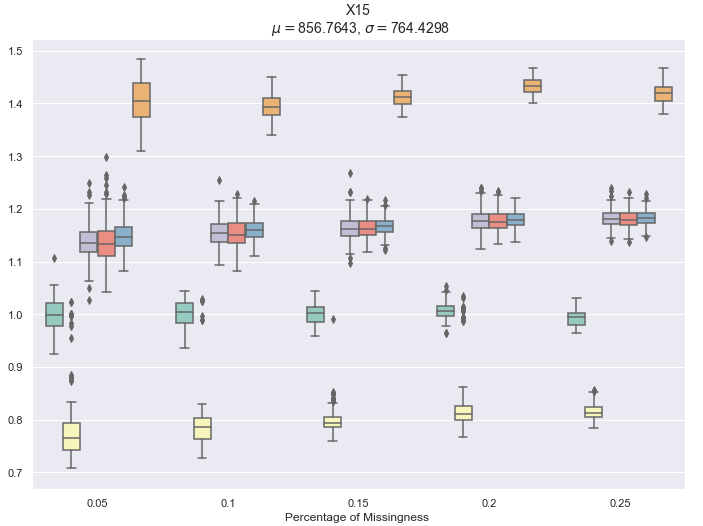}}
		}%
		\subfigure[Performance based on RMSE: X16]{%
			\label{fig:second}
			\frame{\includegraphics[width=0.33\textwidth]{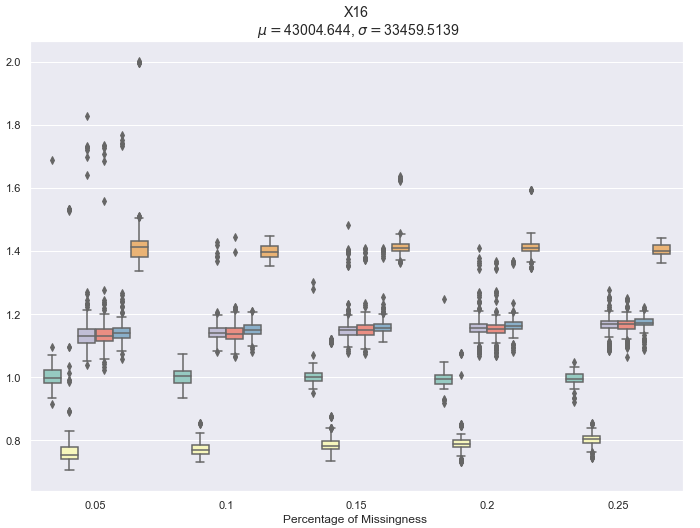}}
		}%
		\subfigure[Performance based on RMSE: X17]{%
			\label{fig:fourth}
			\frame{\includegraphics[width=0.33\textwidth]{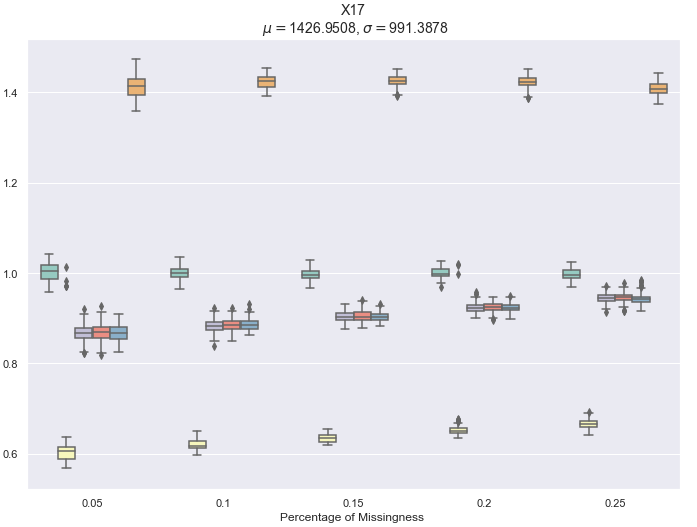}}
		}%
	\\
		\subfigure[Performance based on RMSE: X18]{%
			\label{fig:first}
			\frame{\includegraphics[width=0.33\textwidth]{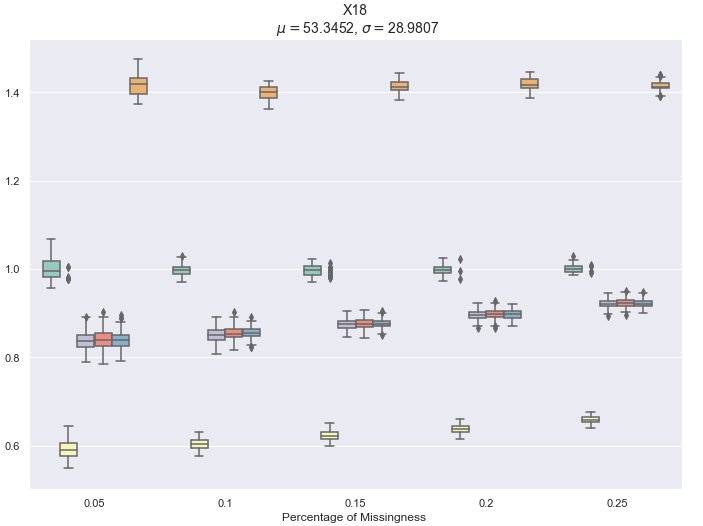}}
		}%
		\subfigure[Performance based on RMSE: X19]{%
			\label{fig:second}
			\frame{\includegraphics[width=0.33\textwidth]{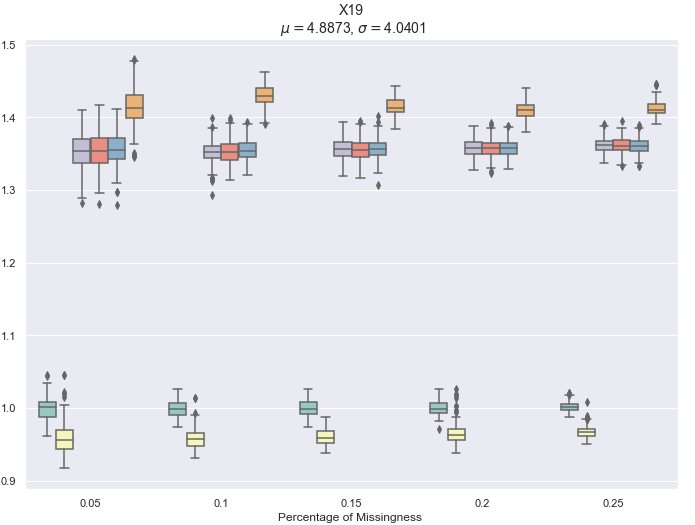}}
		}%
		\subfigure[Performance based on RMSE: X20]{%
			\label{fig:fourth}
			\frame{\includegraphics[width=0.33\textwidth]{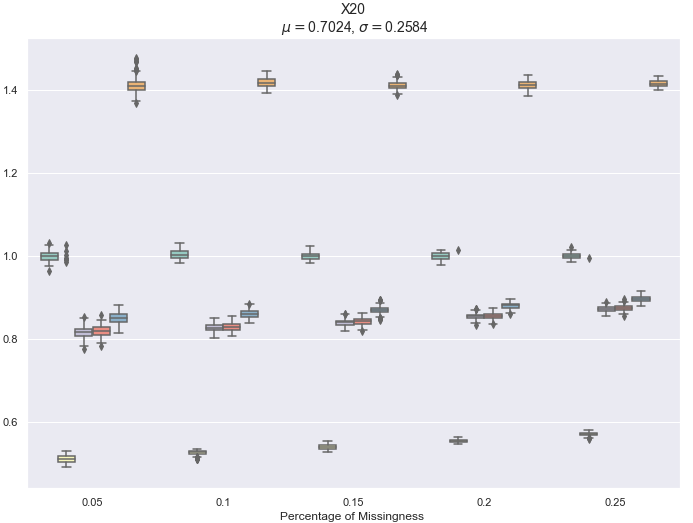}}
		}%
	\end{center}
\end{figure*}


\end{document}